\documentclass[aps,twocolumn,10pt,superscriptaddress,floatfix,nofootinbib,longbibliography]{revtex4-2}
\usepackage[T1]{fontenc}
\usepackage{lmodern}
\usepackage{amsmath,amssymb,amsthm,graphicx,booktabs,bm,bbm}
\usepackage[colorlinks=true,linkcolor=blue,citecolor=blue,urlcolor=blue]{hyperref}
\hypersetup{pdftitle={Measurement-order dependence bounds uncertainty, contextuality, steering, and Bell nonlocality},pdfauthor={Shengjun Wu}}
\newtheorem{theorem}{Theorem}
\newtheorem{proposition}{Proposition}
\newtheorem{lemma}{Lemma}
\newtheorem{corollary}{Corollary}
\newtheorem{conjecture}{Conjecture}
\theoremstyle{definition}
\newtheorem{definition}{Definition}
\theoremstyle{remark}
\newtheorem{remark}{Remark}
\newcommand{\Tr}{\operatorname{Tr}}
\newcommand{\TV}{\operatorname{TV}}
\newcommand{\Var}{\operatorname{Var}}
\newcommand{\ran}{\operatorname{ran}}
\newcommand{\Djs}{D_{\mathrm{JS}}}
\newcommand{\cJ}{\mathcal J}
\newcommand{\cT}{\mathcal T}
\newcommand{\cH}{\mathcal H}
\newcommand{\one}{\mathbbm{1}}
\newcommand{\ket}[1]{|#1\rangle}
\newcommand{\proj}[1]{|#1\rangle\langle#1|}
\newcommand{\Ustar}{U_*}
\newcommand{\qmu}{q_{\mathrm{MU}}}
\newcommand{\bC}{\beta_{\mathrm C}}
\newcommand{\bQ}{\beta_{\mathrm Q}}
\allowdisplaybreaks
\begin{document}
\title{Measurement-order dependence bounds uncertainty, contextuality,\\
steering, and Bell nonlocality}
\author{Shengjun Wu}
\email{sjwu@nju.edu.cn}
\affiliation{National Laboratory of Solid State Microstructures and School of Physics, Collaborative Innovation Center of Advanced Microstructures, Nanjing University, Nanjing 210093, China}

\date{November 18, 2026}

\begin{abstract}
Two quantum measurements applied to the same input in opposite orders give
two joint distributions of the same outcomes. For L\"uders instruments we
measure how far apart these order records can be, over all inputs, by their
total-variation distance $\cT$, their Jensen--Shannon divergence $\cJ$ (in
bits), and their squared Hellinger distance. These order scores vanish
exactly when the measurements commute, and $\cT=\|[A,B]\|/4$ for two binary
sharp observables. Some pairs of bases exceed mutually unbiased bases in
squared Hellinger score, and in odd dimensions in $\cJ$, although their
minimal entropy sum is smaller. For projective measurements, a Maassen--Uffink bound $q>0$ implies
$\cT\ge1-2^{-q}$ and a positive lower bound on $\cJ$. Measuring all
observables of a scenario in sequence defines a joint distribution, so
violations of noncontextuality, steering, and Bell inequalities are bounded
by distances between order records. For correlation Bell inequalities with
binary sharp observables, the squared Bell operator is a symmetrized (Jordan) part minus tensor products
of local commutators. If one party has two settings, the
violation is bounded bilinearly in the two parties' order scores; we
conjecture this bound in general. On Clifford realizations it would follow
from a Hilbert--Schmidt variant of Grothendieck's inequality, which we prove
when one party has at most three settings. For three binary sharp
measurements, the violation of Bell's original inequality by a maximally
entangled state equals an order effect in three-measurement records of the
maximally mixed state minus four times an alternation probability. For the
rank-one Klyachko--Can--Binicio\u{g}lu--Shumovsky pentagon on a qutrit, the
violation is at most the largest product of the order scores of two crossing
diagonals, with equality at the maximal violation.
\end{abstract}
\maketitle

\section{Introduction}\label{sec:intro}
If two classical observables can be read without disturbing the system, the
order of reading does not change their joint statistics. A quantum measurement is implemented by an instrument, which specifies for
each outcome both its probability and the state left behind. Because the first
measurement changes the state, the order matters. Applying $A$ and then $B$ to an input $\rho$ gives a joint
distribution of the two outcomes that in general differs from the one
obtained by applying $B$ first. We call these distributions \emph{order records}. In the experiment we consider, a classical coin selects the order and both
outcomes are recorded. The task is to infer the order from the outcomes. An
\emph{order score} is the largest statistical distance between the two records
over all inputs, and it quantifies how well this task can be performed.

Mutually commuting sharp observables admit a joint distribution and hence
no Kochen--Specker contextuality for their joint projective
contexts~\cite{Budroni22,Fine82}. More generally, jointly measurable settings
(marginals of a single parent POVM) on the steering party cannot demonstrate
steering, and jointly measurable settings on either party preclude bipartite
Bell nonlocality~\cite{Heinosaari16,Guhne23RMP,Wolf09,Quintino14,Uola14,Uola15}.
For POVMs, joint measurability of the individual measurements alone does not
make arbitrary choices of context-specific joint measurements
noncontextual. Our contextuality bounds specify the joint distributions
through commuting effects and their L\"uders instruments.
Two commuting sharp (projective) observables also have a common eigenvector,
so their state-independent uncertainty bounds vanish. We refer to uncertainty,
contextuality, steering, and Bell nonlocality as \emph{signatures} of
incompatibility. Unless stated otherwise, we use the L\"uders instrument, the state update
$\rho\mapsto\sqrt E\rho\sqrt E$ for an outcome with effect $E$, the operator
that gives its probability $\Tr(\rho E)$. For a sharp measurement this is the
projection postulate. For L\"uders instruments, order independence on all inputs is equivalent to
commutativity (Proposition~\ref{prop:basic}).

Our results turn these qualitative facts into inequalities: a signature of a given strength requires a definite amount of order
dependence. Order dependence alone does not guarantee a signature for a
specified state and scenario, or positive state-independent uncertainty.
This does not exclude the familiar existential converse: every
noncommuting binary sharp pair violates CHSH with a suitable state and
partner (Corollary~\ref{cor:minimal}). Order scores
depend on the instrument and not only on the effects, and for
positive-operator-valued measures (POVMs) a positive order score does not
imply incompatibility (Sec.~\ref{sec:more}).

The uncertainty results below concern projective measurements. They do not
assert that intrinsic outcome randomness of a POVM requires order dependence.

Joint-measurability quantifiers such as incompatibility robustness (the amount
of noise needed to make the measurements jointly measurable) give sharper
criteria for steering, in some cases necessary and sufficient
ones~\cite{Uola15,Cavalcanti16,Chen16}. Order records have two merits of their own: they are measured directly in a
sequential experiment, and for binary sharp observables the total-variation score is a commutator
norm, which bounds the commutator part of the squared Bell operator
(Sec.~\ref{sec:product}).

Order-based quantities have been studied before. In quantum cognition the difference $\Tr\rho(PQP-QPQ)$ between the two orders
of a pair of questions, for a given state and a single pair of answers, is an indicator of noncommutativity~\cite{Atmanspacher12,Wang14}. Gao et al.~\cite{Gao23} control the order coherently with a quantum switch (a
device that applies two operations in a superposition of orders) and estimate a mutual eigenspace disturbance. For rank-one projective
measurements its square equals the uniform average of the total-variation distance
$\TV(p^\psi_{AB},p^\psi_{BA})$ over the eigenstates $\psi$ of one measurement,
whereas we choose the order classically and maximize over inputs. Storrer et al.~\cite{Storrer26} quantify context incompatibility by comparing
one marginal with and without an intervening nonselective measurement, for
a given state. The optimized disturbance of one marginal was studied in
Ref.~\cite{Mandayam14}. A different operational measure uses the accessible
fidelity of an ensemble of measurement eigenstates~\cite{Bandyopadhyay13}.
Our scores compare the full records of
the two orders.

\emph{Summary of results.} Section~\ref{sec:records} defines
three order scores and derives their basic properties. For two binary sharp
observables the total-variation score equals $\|[A,B]\|/4$
(Theorem~\ref{thm:binary}), the commutator norm that also controls the Clauser--Horne--Shimony--Holt
(CHSH) operator~\cite{CHSH69,Landau87,Kiukas10}. For such observables
our bounds are therefore commutator bounds with an operational reading. For projective measurements we bound the affinity (Bhattacharyya coefficient)
of the two records from below, which limits how well the order can be
identified, and we characterize equality (Theorem~\ref{thm:support}).
In every dimension $n\ge3$ we construct pairs of bases with a larger squared
Hellinger score than mutually unbiased bases (MUBs), and for odd $n$ a
larger Jensen--Shannon score, but with a strictly smaller minimal entropy sum
(Theorem~\ref{thm:separation}). MUBs maximize the minimal entropy sum, the
optimal constant of the entropic uncertainty relation. The order scores
therefore rank pairs of measurements differently from entropic uncertainty. The optimized disturbance of a single marginal never exceeds the
total-variation score, and it can be strictly smaller, although only slightly
in our example (Remark~\ref{rem:marginal}).

Section~\ref{sec:uncert} shows that entropic and variance uncertainty
relations give lower bounds on the order scores. The arguments are short, and
Fig.~\ref{fig:minimal}(b) shows that the bounds are tight only for the most
uncertain pairs. Section~\ref{sec:sequential} uses the fact that a sequence of measurements is
one joint measurement~\cite{Heinosaari16,HeinosaariWolf10}, so that the
violation of a noncontextuality, steering, or Bell inequality is bounded by
distances between order records. Related bounds with disturbance terms
underlie sequential tests of contextuality that remain valid under imperfect
compatibility~\cite{Guhne10,Szangolies13,Kujala15}. Our inequalities are
instead statements within quantum theory, and they identify which order
records must differ.

The steering and Bell bounds of Sec.~\ref{sec:sequential} are linear in the
order scores of one party. Section~\ref{sec:product} derives bounds that are bilinear in the order
scores of the two parties. For correlation
Bell inequalities with binary sharp observables, the squared Bell operator
equals a Jordan part, which contains only symmetrized products of the local
observables, minus a sum of tensor products of local commutators
(Theorem~\ref{thm:product}). The Jordan part is at most the squared classical bound whenever one party's
observables admit an operator cut decomposition, which writes their
symmetrized products as mixtures of products of classical signs. This is the
case whenever that party has two settings. The squared quantum value then exceeds the squared
classical bound by at most a term bilinear in the two parties' order scores,
with equality for CHSH (Landau's identity). In general the Jordan part can be
larger, but we conjecture that the bilinear bound still holds
(Conjecture~\ref{conj:product}). On Clifford realizations (observables built
from anticommuting generators) the conjecture would follow from a
Hilbert--Schmidt variant of Grothendieck's inequality~\cite{Krivine79,Briet14}
(Proposition~\ref{prop:HSG}). Grothendieck's inequality bounds the quantum
value of every correlation functional by a universal constant times its
classical bound~\cite{Tsirelson87}, and $K_G(r)$ denotes the best constant
when the correlations are inner products of unit vectors in $\mathbb R^r$. We prove the Hilbert--Schmidt variant when one party has at most three
settings. If it holds for all numbers of settings, it implies Krivine's bound
$K_G(2)\le\sqrt2$. Theorem~\ref{thm:steer} gives the analogous steering bound for
pairwise anticommuting trusted measurements.

Section~\ref{sec:three} gives bounds of the same form for two further
scenarios. Two-measurement records of the maximally mixed state do not
depend on the order, but three-measurement records do. The violation of
Bell's original inequality~\cite{Bell64} by a maximally entangled state
equals an order-dependent correlation of these records minus four times the probability that the three recorded outcomes alternate in
sign (Theorem~\ref{thm:three}). For the rank-one pentagon of Klyachko, Can, Binicio\u{g}lu, and Shumovsky
(KCBS) on a qutrit~\cite{KCBS08}, the violation is at most the largest
product of the order scores of two incompatible pairs that form crossing
diagonals of the pentagon, with equality at the maximal violation
(Theorem~\ref{thm:kcbs}). We conjecture this bound for projectors of any rank in every dimension.
Section~\ref{sec:more} delimits what order scores detect for more than two
measurements and for unsharp measurements, and Table~\ref{tab:summary} lists
the bounds. Proofs are collected in the appendices.

\section{Order records and order scores}\label{sec:records}
\subsection{Definitions}\label{sec:defs}
In this article, all Hilbert spaces are finite dimensional and written $\mathbb C^d$, and
$\|\cdot\|$ is the operator norm. Measurement $i$ is described by an instrument $\{\mathcal M^{(i)}_x\}_x$, a
family of completely positive maps whose sum preserves the trace. We take
$\mathcal M^{(i)}_x(\rho)=K^{(i)}_x\rho K^{(i)\dagger}_x$, with one Kraus
operator per outcome. Its effects are $E^{(i)}_x=\mathcal
M^{(i)*}_x(\one)=K^{(i)\dagger}_xK^{(i)}_x$, where $\mathcal M^*$ denotes the
dual (Heisenberg-picture) map, $\Tr[\mathcal M(\rho)X]=\Tr[\rho\,\mathcal
M^*(X)]$. Outcome $x$ occurs with probability $\Tr\mathcal
M^{(i)}_x(\rho)=\Tr(\rho E^{(i)}_x)$ and leaves the system in the state
$\mathcal M^{(i)}_x(\rho)$ up to normalization. The effects fix the statistics
of a measurement performed alone. The instrument also fixes how the
measurement disturbs later ones.

Unless stated otherwise, the instrument considered in this article is the
L\"uders instrument $\mathcal
M^{(i)}_x(\rho)=\sqrt{E^{(i)}_x}\rho\sqrt{E^{(i)}_x}$. For a projective
measurement this is the projection postulate $\rho\mapsto P_x\rho P_x$. We use it because its order dependence is equivalent to noncommutativity. For
L\"uders instruments the order records coincide on all inputs exactly when the
effects commute [Proposition~\ref{prop:basic}(iii)]. Other instruments with
the same effects can produce order dependence of purely classical origin (see
the example at the end of this subsection).

A \emph{schedule} $\pi=(\pi_1,\ldots,\pi_m)$ is an ordering of the $m$
measurements. In one run they are performed one after the other on the same
system, each on the state left by the previous ones. The input $\rho$ is
prepared before the schedule is drawn, the outcomes are labeled by measurement
rather than by time, and the final state is discarded. The first convention
excludes inputs that depend on the order. The second puts the records of all schedules on the same set of strings
$(x_1,\ldots,x_m)$, where $x_i$ is the outcome of measurement $i$ whenever it
is performed. Two records can then be compared entry by entry.
The \emph{order record} of $\pi$ is
\begin{equation}
 p^\rho_\pi(x_1,\ldots,x_m)=\Tr\big[\mathcal M^{(\pi_m)}_{x_{\pi_m}}\circ
 \cdots\circ\mathcal M^{(\pi_1)}_{x_{\pi_1}}(\rho)\big].
  \label{eq:record}
\end{equation}
Measurement $\pi_1$ is performed first and $\pi_m$ last.
For two measurements $A$ and $B$ with instruments $\{\mathcal A_a\}$ and
$\{\mathcal B_b\}$ and effects $E_a$ and $F_b$ we write
$p^\rho_{AB}(a,b)=\Tr(\rho X_{ab})$ and $p^\rho_{BA}(a,b)=\Tr(\rho Y_{ab})$,
where $X_{ab}=\mathcal A_a^*(F_b)$ and $Y_{ab}=\mathcal B_b^*(E_a)$. For L\"uders instruments $X_{ab}=\sqrt{E_a}F_b\sqrt{E_a}$ and $Y_{ab}=\sqrt{F_b}E_a\sqrt{F_b}$. The operators $X_{ab}$ and $Y_{ab}$ are the effects of the sequential measurements ``$A$ then $B$'' and ``$B$ then $A$''. Each family is a POVM on the pairs $(a,b)$, and the order scores defined below are distances between these two POVMs, maximized over inputs.

If the measurements did not disturb one another, every schedule would give the same record. We compare the two records of a pair by three standard measures of the
distance between probability distributions.

\begin{definition}[Order scores]\label{def:scores}
For two measurements and an input $\rho$ let
$\cT(A,B|\rho)=\TV(p^\rho_{AB},p^\rho_{BA})$, where
$\TV(p,q)=\frac12\sum_i|p_i-q_i|$ is the total-variation distance,
$\cJ(A,B|\rho)=\Djs(p^\rho_{AB},p^\rho_{BA})$, where
$\Djs(p,q)=H\big(\frac{p+q}2\big)-\frac12[H(p)+H(q)]$ is the Jensen--Shannon
divergence~\cite{Lin91} and $H$ is the Shannon entropy in bits, and
$\cH^2(A,B|\rho)=1-\sum_{ab}[p^\rho_{AB}(a,b)p^\rho_{BA}(a,b)]^{1/2}$, the
squared Hellinger distance. The \emph{order scores} $\cT(A,B)$, $\cJ(A,B)$,
and $\cH^2(A,B)$ are their maxima over $\rho$. The Jensen--Shannon divergence
extends directly to more than two schedules, which is needed for scenarios
with several measurements (Secs.~\ref{sec:sequential} and~\ref{sec:more}).
For a set $\Sigma$ of schedules drawn uniformly, the \emph{order information}
of $\rho$ is the mutual information between the schedule and the record,
\begin{equation}
 \cJ_\Sigma(\rho)=H(\bar p^\rho)-\frac1{|\Sigma|}\sum_{\pi\in\Sigma}
 H(p^\rho_\pi)\in[0,\log_2|\Sigma|],
 \label{eq:orderinfo}
\end{equation}
where $\bar p^\rho$ is the average record. For the two schedules of a pair, $\cJ_\Sigma(\rho)=\cJ(A,B|\rho)$.
\end{definition}

The three scores answer different questions. If the order is chosen by a fair coin, the optimal probability of guessing it
from one record is $(1+\cT)/2$. The score $\cJ$ is the largest number of bits about the order that one record
provides. The quantity
$1-\cH^2=\sum_{ab}(p_{AB}\,p_{BA})^{1/2}$ is the Bhattacharyya coefficient, or
affinity, of the two records. It controls repeated trials: if the same order, chosen by a fair coin, is
applied to $N$ copies of an input $\rho$, the optimal guess from the $N$
records is wrong with probability at most $\frac12[1-\cH^2(A,B|\rho)]^N$. We call a difference between two order records of the same input an
\emph{order effect}, and we measure its size by a distance between the
records.

The total-variation distance is the largest difference between the
probabilities that the two distributions assign to an event,
$\TV(p,q)=\max_S[p(S)-q(S)]$. Maximizing over inputs gives
\begin{equation}
 \cT(A,B)=\max_{S}\lambda_{\max}\Big(\sum_{(a,b)\in S}(X_{ab}-Y_{ab})\Big),
 \label{eq:Tspectral}
\end{equation}
where $S$ ranges over sets of outcome pairs. The score $\cT$ is therefore
obtained exactly from finitely many eigenvalue problems, one for each $S$.
Not every divergence gives a finite score. The relative
entropy of the two records, for instance, is infinite on some input for every
noncommuting pair of projective measurements (Appendix~\ref{app:basic}).

Order scores belong to instruments, not to effects. Consider two measurements of the same qubit observable, with effects $\proj0$
and $\proj1$. Measurement $A$ uses the L\"uders instrument, and measurement
$B$ applies $\sigma_x$ to the state after reading the outcome. On the input
$\ket0$ the order $AB$ gives $(a,b)=(0,0)$ and the order $BA$ gives
$(a,b)=(1,0)$. The two records are disjoint, so $\cJ=1$ bit, whereas two L\"uders instruments with these effects
give $\cJ=0$. A classical device that disturbs what it reads behaves in the
same way. A positive order score is therefore a statement about the specified
instruments and not a device-independent witness of nonclassicality.

\subsection{Basic properties}\label{sec:basic}
The following proposition collects the properties used throughout. Part (i) shows that the three scores bound one another, so that a bound on one
of them gives bounds on the others.

\begin{proposition}[Basic properties]\label{prop:basic}
(i) $0\le\cJ\le1$ bit, and pointwise in $\rho$ (and hence for the maxima over
$\rho$)
\begin{gather}
 1-h\Big(\frac{1+\cT}2\Big)\le\cJ\le\cT, \label{eq:JT}\\
 \cH^2\le\cJ\le\cH^2/\ln2, \label{eq:JH}
\end{gather}
where $h$ is the binary entropy. The constants in Eq.~\eqref{eq:JH} are sharp.
(ii) The maxima are attained on pure states. The scores are invariant
under a common unitary and under relabeling of outcomes, and they cannot
increase under classical postprocessing of the records.
(iii) For L\"uders instruments and the set $\Sigma$ of all $m!$ schedules,
$\cJ_{\Sigma}(\rho)=0$ for every $\rho$ if and only if effects of different
measurements commute. For two measurements the same holds for $\cT$ and
$\cH^2$.
(iv) For a fixed input the records can coincide without commutativity; the
maximally mixed state gives $p_{AB}=p_{BA}$ for every L\"uders pair.
\end{proposition}

Part (ii) concerns postprocessing of the \emph{records}. Coarse-graining a
\emph{measurement}, that is, merging outcomes and using the L\"uders
instrument of the merged effect, is a different operation, because it keeps
coherences between the merged outcomes. Monotonicity under this operation is
a basic requirement for incompatibility
quantifiers~\cite{Heinosaari15,Designolle19}, and we have not established it
for the order scores. Part (iii) rests on the sequential-product criterion of
Gudder and Nagy~\cite{Gudder01}: if $\sqrt EF\sqrt E=\sqrt FE\sqrt F$, then
$E$ and $F$ commute. For L\"uders instruments the order scores are thus
faithful indicators of noncommutativity. For unsharp measurements
noncommuting effects can still be jointly measurable (Sec.~\ref{sec:more}).
Part (iv) is the reason
for maximizing over inputs: a single input, even one of full rank, can hide
the order dependence. It also explains why order effects on the maximally
mixed state require at least three measurements (Sec.~\ref{sec:three}).

\subsection{Two binary sharp observables}\label{sec:binary}
Binary sharp observables are the measurements of the CHSH and other
correlation Bell inequalities (Secs.~\ref{sec:product} and~\ref{sec:three}),
and for them the total-variation score has a closed form. Such an observable
is a Hermitian involution, $A=A^\dagger$ and $A^2=\one$, with eigenvalues
$\pm1$. Spin components of a qubit and parities are examples. Let $A$ and $B$
be Hermitian involutions with spectral projectors $P_\pm=(\one\pm A)/2$ and
$Q_\pm=(\one\pm B)/2$. By Jordan's lemma~\cite{Halmos69}, $\mathbb C^d$ splits
into subspaces of dimension one or two that are invariant under both $A$ and
$B$, which we call blocks. The pair is thus a direct sum of qubit pairs and of
commuting one-dimensional pairs. On the $j$th two-dimensional block, $A=\sigma_z$ and
$B=\cos\theta_j\,\sigma_z+\sin\theta_j\,\sigma_x$ in a suitable basis, with
$\theta_j\in(0,\pi)$ the angle between the Bloch vectors; on a
one-dimensional block $\theta_j\in\{0,\pi\}$.

\begin{theorem}[Binary sharp observables]\label{thm:binary}
For L\"uders instruments,
\begin{equation}
 \cT(A,B)=\tfrac14\|[A,B]\|=\max_j\tfrac12\sin\theta_j .
 \label{eq:binaryT}
\end{equation}
Let $D$ be a divergence between probability distributions that is jointly
convex, invariant under relabeling of the outcomes, and zero on equal
arguments. Then $\max_\rho D(p^\rho_{AB},p^\rho_{BA})=\max_j\Psi_D(\tfrac12\sin
\theta_j)$, where $\Psi_D(t)$ denotes the maximum of
$D(p^\rho_{AB},p^\rho_{BA})$ over the inputs of a single qubit pair with
$\frac12\sin\theta=t$. For $D=\Djs$ we write $\Psi=\Psi_{\rm JS}$; then
$\Psi(1/2)=h(3/4)-\frac12\approx0.3113$ bits and
$t^2/(2\ln2)\le\Psi(t)\le t$.
\end{theorem}

The theorem gives the commutator norm of two binary sharp observables an
operational meaning: the order of the two measurements can be guessed from
one record with probability at most $(1+\|[A,B]\|/4)/2$. Every order score of
a binary sharp pair is a function of its block angles alone. For $\cT$, the block closest to anticommuting decides, and $\cT\le1/2$ with equality exactly
when $A$ and $B$ anticommute on some block. Even for an anticommuting qubit
pair one record identifies the order only with probability $3/4$ and carries
$\Psi(1/2)\approx0.31$ bits.

On a grid of $250$ points $\Psi$ is strictly increasing
[Fig.~\ref{fig:minimal}(a)], but we have no proof of monotonicity. If $\Psi$
is increasing, then $\cJ(A,B)=\Psi(\cT(A,B))$ for every binary sharp pair. The
optimal inputs for $\cT$ and $\cJ$ differ. For a qubit pair with $\theta=0.6$ and Bloch vectors $\bm a$ and $\bm b$ of
$A$ and $B$, the total-variation distance is largest on an eigenstate of the
observable along $\bm a-(\bm a\cdot\bm b)\bm b$. The Jensen--Shannon optimum ($\Psi\approx0.1119$ bits) has Bloch
vector $(-0.074,0,0.997)$ or its image $(0.624,0,0.781)$ under
$A\leftrightarrow B$, close to an eigenvector of $A$ or of $B$.

\subsection{Two projective measurements}\label{sec:pvm}
We now turn to projective measurements with any number of outcomes and ask
two questions. How distinguishable can the two order records of an input be?
Which pairs of measurements have the largest order scores? For the second
question the expected answer is mutually unbiased bases (bases with $|\langle
a|b\rangle|^2=1/d$ for all $a,b$), which are the most uncertain pairs
according to the entropic uncertainty relation.
Theorem~\ref{thm:support} answers the first question.
Theorem~\ref{thm:separation} shows that the expected answer to the second is
wrong for $\cH^2$ in every dimension $n\ge3$ and for $\cJ$ in every odd
dimension.

Let $A=\{P_a\}_{a=1}^n$ and $B=\{Q_b\}_{b=1}^m$ be projective measurements
(PVMs) of arbitrary ranks. For a unit vector $\psi$ let $p_a=\|P_a\psi\|^2$
and $q_b=\|Q_b\psi\|^2$ be the probabilities of the outcomes of $A$ and of $B$
on $\psi$. We call $a$ and $b$ occupied if $p_a>0$ and $q_b>0$, and write
$r_\psi$ and $s_\psi$ for the numbers of occupied $a$ and $b$. These numbers
matter because the record of the order $AB$ vanishes unless $a$ is occupied
and that of $BA$ vanishes unless $b$ is occupied, so the two records can
overlap only on the $r_\psi\times s_\psi$ rectangle of occupied pairs.

To quantify how well the order can be identified from many records we use
the Chernoff coefficients
\[
 \chi_z(\psi)=\sum_{ab}\big(p^\psi_{AB}(a,b)\big)^z\,
 \big(p^\psi_{BA}(a,b)\big)^{1-z},\qquad 0<z<1.
\]
They satisfy $\chi_z\le1$, with equality only if the two records coincide,
and $\chi_z=0$ if the records have disjoint supports. Smaller values mean
more distinguishable records. At $z=1/2$ the coefficient is the affinity,
$\chi_{1/2}(\psi)=1-\cH^2(A,B|\psi)$. The whole family is needed for the
asymptotic error rate. If a fixed unknown order is applied to $N$ independent
copies of the input $\psi$, the optimal probability of misidentifying the
order decays as $2^{-N\xi(\psi)}$ with the Chernoff exponent
$\xi(\psi)=-\log_2\inf_{0<z<1}\chi_z(\psi)$~\cite{Chernoff52}. A lower bound
on $\chi_z$ is therefore an upper bound on the distinguishability of the two
orders.

\begin{theorem}[Affinity bound]\label{thm:support}
For every unit vector $\psi$ and $0<z<1$,
\begin{equation}
 \chi_z(\psi)\ge\Big[\Big(\sum_a p_a^{1-z}\Big)\Big(\sum_b q_b^z\Big)\Big]^{-1}
 \ge r_\psi^{-z}s_\psi^{z-1}.
 \label{eq:support}
\end{equation}
Consequently $\cH^2(A,B|\psi)\le1-(r_\psi s_\psi)^{-1/2}$,
$\cH^2(A,B)\le1-(nm)^{-1/2}$, and $\xi(\psi)\le\log_2\max(n,m)$.
For $n,m\ge2$ the last bound is sharp: after interchanging the measurements
if necessary, the infimum is approached as $z\to0$.
If either PVM has one outcome, both records coincide and $\xi=0$.
Equality $\chi_z(\psi)=r_\psi^{-z}s_\psi^{z-1}$ holds if and only if $p$ and $q$ are uniform
on their supports and $P_aQ_b\psi=P_a\psi/s_\psi$ and $Q_bP_a\psi=Q_b\psi/r_\psi$
for all occupied $a,b$. For given numbers $r$ and $s$ of occupied outcomes,
PVMs and a vector $\psi$ that satisfy these conditions exist if and only if
$d\ge r+s-1$.
\end{theorem}

The theorem says that the two records of a projective pair always have a
positive affinity. The reason is that the inner products
$\langle P_a\psi,Q_b\psi\rangle$ sum to
$\langle\psi,\psi\rangle=1$, and each of them is small unless both records
have weight on the pair $(a,b)$. Consequently the order of two L\"uders
projective measurements can never be identified with certainty from one
record: $\cH^2<1$, $\cT<1$, and $\cJ<1$ bit, in contrast to the instrument
example of Sec.~\ref{sec:defs}. The fewer outcomes the input occupies, the
stronger the constraint. The bound on $\xi$ means that the error probability
with $N$ copies cannot decay at a faster exponential rate than
$\max(n,m)^{-N}$. The equality
condition means that the normalized postmeasurement vectors (branches)
$\sqrt{r_\psi}P_a\psi$ and $\sqrt{s_\psi}Q_b\psi$ span two subspaces that meet
exactly in the line through $\psi$ and are orthogonal otherwise.

We compare the order scores with a standard state-independent measure of the
uncertainty of two projective measurements, the minimal entropy sum
$\Ustar(A,B)=\min_\rho[H(A)_\rho+H(B)_\rho]$, where $H(A)_\rho$ is the
Shannon entropy of the outcome distribution $\Tr(\rho P_a)$. It is the
optimal constant in the entropic uncertainty relation
$H(A)_\rho+H(B)_\rho\ge\Ustar$~\cite{Coles17}. For two bases of $\mathbb C^d$
it is at most $\log_2d$, and it attains this value for MUBs~\cite{MU88}. The next theorem shows that
the pairs that are extremal for $\Ustar$ need not be extremal for the order
scores.

Theorem~\ref{thm:separation} below has three parts. Part (a) treats the
benchmark pairs. For MUBs, an eigenstate of one basis
gives the uniform distribution over the outcomes of the other. The same
property for projectors of arbitrary rank reads $P_aQ_bP_a=P_a/n$ and
$Q_bP_aQ_b=Q_b/n$: the first condition says that every state in the range of
$P_a$ gives each outcome $b$ with probability $1/n$, and the second says the
same with the roles exchanged. Such pairs are called mutually unbiased
measurements~\cite{Tavakoli21,Farkas23}. The conditions imply that both
measurements have $n$ outcomes and that all projectors have the same rank,
and for rank one they define MUBs in dimension $n$. After one of the two
measurements the outcome of the other is uniform for every input, so the
records are $p_{AB}(a,b)=p_a/n$ and $p_{BA}(a,b)=q_b/n$, and all scores can be
computed.

Part (b) realizes the equality case of Theorem~\ref{thm:support} for an input
that occupies $r$ outcomes of $A$ and $s$ outcomes of $B$. For $k=1$ the two
measurements are bases of $\mathbb C^n$, and $n=r+s-1$ is the smallest
dimension that allows equality. For $k>1$ all projectors have rank $k$, which
shows that the separation is not special to bases.
Part (c) treats direct sums.

\begin{theorem}[Benchmarks and separation]\label{thm:separation}
(a) Let $A$ and $B$ be mutually unbiased measurements, that is,
$P_aQ_bP_a=P_a/n$ and $Q_bP_aQ_b=Q_b/n$ for all $a,b$. Then
$\cH^2=1-n^{-1/2}$, $\cT=1-1/n$,
$\max_\psi\xi(\psi)=\log_2n$, and $\Ustar=\log_2n$. For a pair of MUBs in dimension $d\ge2$ (rank one, $n=d$),
\begin{equation}
 \varphi(1/d)\le\cJ<\frac{d-1}{d+1},\qquad
 \varphi(x)=h\Big(\frac{1+x}2\Big)-\frac{h(x)}2,
 \label{eq:mubJ}
\end{equation}
with equality on the left for $d=2$.
(b) For all integers $n\ge3$, $k\ge1$, and $r,s\ge2$ with $r+s=n+1$ there are
two $n$-outcome PVMs on $\mathbb C^{nk}$, with all projectors of rank $k$,
such that
\begin{equation}
 \begin{aligned}
 \cH^2&\ge1-(rs)^{-1/2}>1-n^{-1/2},\\
 \Ustar&=\log_2\min(r,s)<\log_2n .
 \end{aligned}
 \label{eq:sep}
\end{equation}
For odd $n$ and $r=s$ they also satisfy $\cJ\ge(n-1)/(n+1)$, which exceeds
$\cJ$ of every pair of MUBs in dimension $n$.
(c) Suppose $\mathbb C^d=\bigoplus_j\mathcal K_j$ and every $P_a$ and every $Q_b$
is block diagonal. Then each order score equals its largest value over the
blocks, and $\Ustar$ equals its smallest value over the blocks.
\end{theorem}

The function $\varphi(x)$ is the Jensen--Shannon divergence between the
distribution $(1,0)$ of a sure event and the distribution $(x,1-x)$ of an
event of probability $x$. It decreases from $\varphi(0)=1$ bit to
$\varphi(1)=0$. It appears because on an eigenstate of $A$ the outcome of $A$
is certain in the order $AB$ and recurs only with probability $x$ in the
order $BA$.

For $k=1$, part (b) compares bases in the same dimension. The constructed
pair has a larger $\cH^2$ than any pair of MUBs, and for odd $n$ a larger
$\cJ$, although MUBs have the largest possible $\Ustar$. The reason is
visible in Theorem~\ref{thm:support}. For bases of $\mathbb C^n$ its equality
case requires $r+s\le n+1$, and the affinity $(rs)^{-1/2}$ is then smallest
for $r$ and $s$ close to $(n+1)/2$. For MUBs the second outcome is always
uniform, and the affinity cannot fall below $n^{-1/2}$. With
$r=\lfloor(n+1)/2\rfloor$ and $s=\lceil(n+1)/2\rceil$ the entropy gap $\log_2[n/\min(r,s)]$ tends to one bit as $n\to\infty$, and
the lower bound $n^{-1/2}-(rs)^{-1/2}$ on the gap in squared Hellinger score
is largest, $0.135$, at
$n=11$. The comparison does not extend to $\cT$. For $n=3$ the constructed pair has
$\cT=0.640$ by Eq.~\eqref{eq:Tspectral}, whereas MUBs have $\cT=2/3$. In part (c) the order scores are decided by the most incompatible block and
$\Ustar$ by the most compatible one, so pairs with $\Ustar=0$ can have large
order scores. An example is the computational basis
together with the columns of $1\oplus F_{d-1}$, where $F_t$ is the
$t$-dimensional Fourier matrix with entries $t^{-1/2}e^{2\pi\mathrm i jl/t}$.
The two bases share their first vector, so $\Ustar=0$, and on the remaining
block they are MUBs in dimension $d-1$. This pair has
$\cJ\ge\varphi(1/(d-1))$. For the Fourier MUB pairs with $d=3,4,5$ we find
$\cJ=\varphi(1/d)$ to eight digits, and we conjecture equality on the left of
Eq.~\eqref{eq:mubJ} for all MUBs.

\begin{remark}[Full record versus disturbed marginal]\label{rem:marginal}
Earlier measures of disturbance keep only one outcome. The optimized
disturbance of $A$ by $B$~\cite{Mandayam14} compares the distribution of $A$
measured alone with its distribution when $B$ is measured first, and maximizes
their distance over inputs. We use the total-variation distance. Both
distributions are marginals of the two order records, so this quantity never
exceeds $\cT$. The inequality can be strict. Take
the computational basis of $\mathbb C^3$ and the columns of the reflection
$U=\one-2\bm v\bm v^{\mathsf T}/101$ with $\bm v=(1,6,8)$. Both optimized
marginal disturbances equal $0.31514$, whereas the input
$(39,68,62)/\sqrt{9989}$ already gives $\cT\ge0.31553$. The difference,
$4\times10^{-4}$, is small, and a random search over qutrit bases found none
larger than $7\times10^{-4}$.
\end{remark}

\section{Uncertainty}\label{sec:uncert}
Uncertainty relations have been related to nonlocality and
steering~\cite{OppenheimWehner10,Berta10,Schneeloch13} and to measurement
noise and disturbance~\cite{Buscemi14,BuschLahtiWerner13}. Here we relate them
to order records. If no state has definite
values of both $A$ and $B$, then an eigenstate of $A$ is disturbed by a
measurement of $B$, and the outcome of $A$ is certain in one order and
uncertain in the other. An uncertainty relation therefore gives a lower
bound on the order scores.

Let $A=\{P_a\}_{a=1}^n$ and $B=\{Q_b\}_{b=1}^m$ be PVMs,
$c=\max_{a,b}\|P_aQ_b\|$, and $\qmu=-\log_2c^2$. For two bases, $c$ is the largest overlap $|\langle a|b\rangle|$. In general $c^2$ is the largest
probability of an outcome $b$ on a state in the range of some $P_a$, and
$c=1$ exactly when the ranges of some $P_a$ and $Q_b$ share a nonzero vector. The
Maassen--Uffink relation~\cite{MU88,Krishna02,Coles17} states that
$H(A)_\rho+H(B)_\rho\ge\qmu$ for every $\rho$, so that $\Ustar\ge\qmu$. Let
$k=\min(n,m)\ge2$ and $g_k(x)=h(x)+x\log_2(k-1)$.
If $k=1$, one measurement is the identity PVM and
$\Ustar=\cT=\cJ=0$; no inverse of $g_1$ is needed.
For $k\ge2$, $g_k(x)$ is the largest Shannon
entropy of a distribution on $k$ outcomes in which one outcome has
probability $1-x$; the bound $H\le g_k(x)$ is Fano's
inequality~\cite{Cover06}. The function $g_k$ increases on $[0,1-1/k]$ from
$0$ to $\log_2k$. We use it to convert the overlap $c$ into a bound on
$\Ustar$. Commuting PVMs have a common eigenvector, so $\Ustar>0$ already
implies $\cJ>0$ by Proposition~\ref{prop:basic}(iii).
Theorem~\ref{thm:uncertainty} makes this implication quantitative in four
versions: with the Maassen--Uffink overlap (a), with the optimal entropic
bound $\Ustar$ (b), with variances and the Robertson bound (c), and, in the
opposite direction, as an uncertainty relation for a given input in terms of
its order information (d).

\begin{theorem}[Uncertainty requires order dependence]\label{thm:uncertainty}
For PVMs of arbitrary ranks (with $k\ge2$ in part (b)):
\begin{enumerate}
\item[(a)] $\cT\ge1-c^2=1-2^{-\qmu}$, with equality for MUBs, and
$\cJ\ge\varphi(c^2)=\varphi(2^{-\qmu})$, with equality for qubit MUBs.
\item[(b)] $\Ustar\le g_k(1-c^2)\le g_k(\min\{\cT,1-1/k\})$. Hence
\begin{equation}
 \cT\ge g_k^{-1}(\Ustar),\qquad \cJ\ge\varphi\big(1-g_k^{-1}(\Ustar)\big),
 \label{eq:Uorder}
\end{equation}
where $g_k^{-1}$ is the inverse on $[0,1-1/k]$; it is defined because
$\Ustar\le\log_2k$.
\item[(c)] For binary sharp observables,
$\max_\rho|\langle[A,B]\rangle_\rho|/2=2\cT$, so the Robertson
bound~\cite{Robertson29} $\Delta A\,\Delta B\ge|\langle[A,B]\rangle|/2$ never exceeds $2\cT$. Also
\begin{equation}
 U_{\rm var}:=\min_\rho(\Var A+\Var B)\le1-\sqrt{1-4\cT^2},
 \label{eq:Uvar}
\end{equation}
with equality when $A$ and $B$ have a single two-dimensional block (qubit
pairs).
\item[(d)] For every input $\rho$, with the R\'enyi entropy of order $1/2$,
$H_{1/2}(p)=2\log_2\sum_a\sqrt{p_a}$,
\begin{equation}
 H_{1/2}(A)_\rho+H_{1/2}(B)_\rho\ge-2\log_2\!\big[1-\ln2\,\cJ(A,B|\rho)\big].
 \label{eq:sameprobe}
\end{equation}
\end{enumerate}
\end{theorem}

Parts (a) and (b) use an eigenvector of $P_aQ_bP_a$ with eigenvalue $c^2$. On
this input the outcome of $A$ is certain, but after a measurement
of $B$ it recurs with probability at most $c^2$. Part (b) replaces the overlap $c$ by the optimal bound $\Ustar$. On the same input $H(A)=0$, and one outcome of $B$ has probability $c^2$, so
Fano's inequality gives $\Ustar\le H(B)\le g_m(1-c^2)$. Exchanging the roles
of $A$ and $B$ gives $\Ustar\le g_n(1-c^2)$, and the smaller of the two bounds
is $g_k(1-c^2)$. A large
$\Ustar$ therefore forces a small $c$ and, by part (a), a large $\cT$. The
number of outcomes $k$ enters through Fano's inequality.

The first statement of part (c) says that the right-hand side of the
Robertson relation is at most $2\cT$ on every state, and that some state
attains this value. In the second statement, $U_{\rm var}$ is the variance
analog of $\Ustar$, the optimal constant in $\Var A+\Var
B\ge U_{\rm var}$. Equation~\eqref{eq:Uvar} is equivalent to
$\cT\ge\frac12[U_{\rm var}(2-U_{\rm var})]^{1/2}$, a lower bound on the order
score in terms of the variance uncertainty. It restates the qubit uncertainty
region~\cite{Abbott16} in terms of $\cT$. In higher dimensions, $U_{\rm var}$ is governed by the block closest to
commuting and $\cT$ by the block closest to anticommuting, so $U_{\rm var}$
can vanish while $\cT=1/2$, as for $A=\sigma_z\oplus1$ and
$B=\sigma_x\oplus1$.

Part (d) goes in the opposite direction and is state dependent. An input whose
two order records differ cannot have definite values of both $A$ and $B$: the
sum of the two R\'enyi entropies is bounded from below by an increasing
function of its order information. The order $1/2$ arises because the proof
uses Theorem~\ref{thm:support} with $z=1/2$. The bound concerns the
\emph{sum}, so one of the two entropies may vanish, as on an eigenstate of
$A$. For a qubit MUB pair, an eigenstate of $A$ has $H_{1/2}(A)=0$ and
$H_{1/2}(B)=1$ bit, and the right-hand side is $0.70$ bits.

\begin{figure*}[t]
\centering
\includegraphics[width=\textwidth]{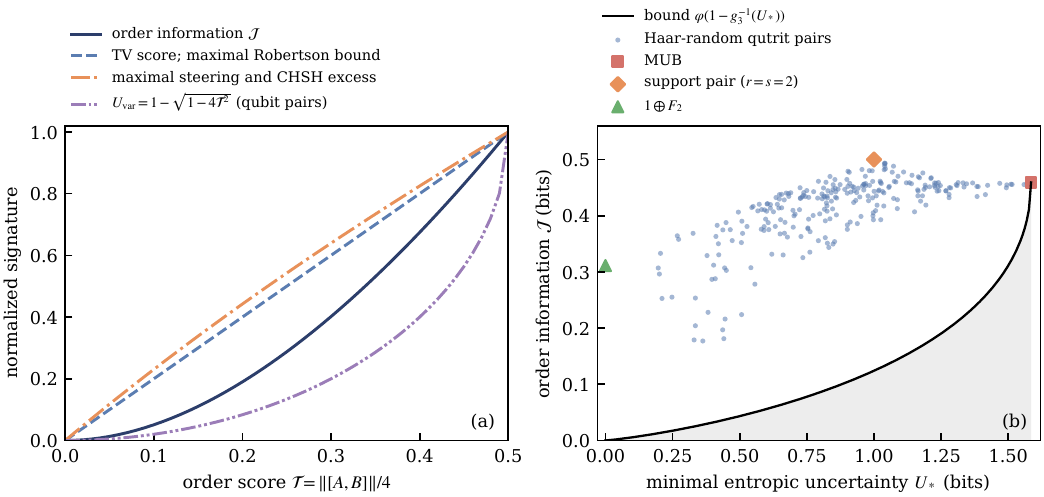}
\caption{(a) Quantities for two binary sharp observables as functions of
$\cT=\|[A,B]\|/4$, each divided by its value at $\cT=1/2$. Shown are the
Jensen--Shannon score $\Psi(\cT)$, which is the largest order information of
a qubit pair (Theorem~\ref{thm:binary}; computed on a grid); $\cT$ itself;
the excess over the classical bound of the maximal steering value
$\sqrt{2+4\cT}$ and of the maximal CHSH value $2\sqrt{1+2\cT}$ with an ideal
partner, whose settings are chosen optimally (Corollary~\ref{cor:minimal};
the two normalized curves coincide); and the right-hand side of
Eq.~\eqref{eq:Uvar}, which equals $U_{\rm var}$ for qubit pairs only. (b) Pairs of bases of $\mathbb C^3$ in the plane of minimal entropy sum $\Ustar$ and Jensen--Shannon score $\cJ$,
the largest order information of the pair. Both are computed by multistart
local optimization, so the plotted $\cJ$ are lower bounds and the plotted
$\Ustar$ upper bounds. The bound~\eqref{eq:Uorder} excludes the shaded region
and is attained numerically by the MUB pair, for which $\cJ=0.459148$, equal
to $\varphi(1/3)$ to this precision. Several random pairs and the pair of
Theorem~\ref{thm:separation}(b) with $r=s=2$ lie above the
MUB value at smaller $\Ustar$, and $1\oplus F_2$ has $\Ustar=0$.}
\label{fig:minimal}
\end{figure*}

Figure~\ref{fig:minimal}(b) shows the qutrit case. The bound~\eqref{eq:Uorder} is attained at the MUB end, where $\Ustar=\log_23$
(numerically for $\cJ$). It is far from tight for generic pairs, because its
proof evaluates the records on an eigenvector of one measurement and not on
the optimal input. Theorem~\ref{thm:separation}(c) shows that no converse bound exists:
pairs with $\Ustar=0$ can have $\cJ$ arbitrarily close to one bit.

\section{Sequential realizations}\label{sec:sequential}
\subsection{A schedule defines a joint distribution}
Contextuality, steering, and Bell nonlocality have a common form: the
statistics of the individual contexts (sets of jointly measurable observables)
or settings cannot be obtained from a single classical model, such as a joint distribution of all
outcomes~\cite{Fine82}. A sequential measurement of all observables on one
copy of the input always produces such a joint distribution. This
distribution is correct on a context if the measurements of that context
are performed first. Otherwise it is wrong only as far as the earlier
measurements have disturbed the context, and this disturbance is a distance
between two order records. Lemma~\ref{lem:seq} states this, and the rest of
the section applies it. We write $p|_C$ for the marginal of a distribution
$p$ on a subset $C$ of the measurements.

\begin{lemma}[Sequential realization]\label{lem:seq}
Let $\pi$ be a schedule of measurements $\mathcal M^{(1)},\ldots,\mathcal
M^{(m)}$. (i) For arbitrary instruments, $p^\rho_\pi$ is a joint distribution
of all $m$ outcomes.
(ii) Let $C$ be a set of measurements whose effects mutually commute and
whose instruments are L\"uders, and let $\pi_C$ perform the measurements in
$C$ first and the others afterward in the order of $\pi$. Then the marginal
of $p^\rho_{\pi_C}$ on $C$ is the joint quantum distribution
$p^\rho_C(x_C)=\Tr(\rho\prod_{i\in C}E^{(i)}_{x_i})$.
(iii) Consequently, for every divergence $D$ that obeys data processing
(in particular, does not increase when both arguments are marginalized),
\begin{equation}
 D\big(p^\rho_\pi|_C,\,p^\rho_C\big)=D\big(p^\rho_\pi|_C,\,p^\rho_{\pi_C}|_C\big)
 \le D\big(p^\rho_\pi,\,p^\rho_{\pi_C}\big).
 \label{eq:seq}
\end{equation}
\end{lemma}

In part (ii) the order within $C$ is irrelevant. The left-hand side of
Eq.~\eqref{eq:seq} compares the sequential distribution with the quantum
statistics of the context, and the right-hand side is an order effect between
the schedules $\pi$ and $\pi_C$ on the same input.

\subsection{Contextuality}\label{sec:context}
A contextuality scenario consists of measurements and a family of
\emph{contexts}, which are sets of measurements with mutually commuting
effects. The measurements of a context can be performed jointly, and a
behavior $\{p_C\}$ assigns a distribution to the outcomes of each context. A
noncontextual hidden-variable model assigns an outcome to every measurement
independently of the context in which it is measured. Equivalently, a behavior
is noncontextual if it is the family of marginals of one joint distribution of
all outcomes~\cite{Fine82,Abramsky17}.

Contextuality is witnessed by noncontextuality inequalities, which are linear bounds on the context
statistics that every noncontextual behavior satisfies. We write such an
inequality as $\beta=\sum_C\sum_{x_C}\alpha_C(x_C)p_C(x_C)\le b$, where the
real coefficients $\alpha_C(x_C)$ weight the outcome strings $x_C$ of context
$C$ and $b$ is the noncontextual bound. We require it to hold for \emph{every}
joint distribution of all outcomes, or equivalently for every deterministic
assignment of outcomes. This requirement is needed because the sequential
distribution $p^\rho_\pi$ is a joint distribution that need not obey further
constraints, such as exclusivity, that some inequalities presuppose (see the
KCBS example below). The quantum value $\beta(\rho)$ is obtained with
$p_C=p^\rho_C$, and $\beta(\rho)>b$ certifies contextuality.

The spread $\omega_C=\max\alpha_C-\min\alpha_C$ converts a statistical error
on context $C$ into an error of $\beta$: replacing the distribution on $C$
by one at total-variation distance $t$ changes the term of $C$ by at most
$\omega_Ct$. As a measure that
does not depend on a particular inequality we use $\delta_{\rm
NC}(\rho)=\min_q\max_C\TV(q_C,p^\rho_C)$, where $q$ ranges over noncontextual
behaviors. It is the distance from the quantum behavior to the nearest
noncontextual one, with the error taken on the worst context.

\begin{theorem}[Contextuality requires order effects]\label{thm:context}
Let all instruments be L\"uders. For every input $\rho$ and every schedule
$\pi$ of all measurements,
\begin{equation}
 \beta(\rho)-b\le\sum_C\omega_C\,
 \TV\big(p^\rho_\pi|_C,\,p^\rho_{\pi_C}|_C\big),
 \label{eq:contextbound}
\end{equation}
and $\delta_{\rm NC}(\rho)\le\min_\pi\max_C\TV(p^\rho_\pi|_C,p^\rho_{\pi_C}|_C)$.
In particular, if $\beta(\rho)>b$, then for every schedule $\pi$ some
context $C$ has
$\TV(p^\rho_\pi|_C,p^\rho_{\pi_C}|_C)\ge v:=(\beta(\rho)-b)/\sum_C\omega_C$,
and the two schedules $\pi$ and $\pi_C$ have order information at least
$1-h((1+v)/2)$ on $\rho$.
\end{theorem}

The theorem says that a violation must show up as an order effect on the
same input. If $\beta(\rho)>b$, then for every way of sequencing all the
measurements there is a context whose statistics change when its
measurements are moved to the front, and the change is at least the
violation divided by $\sum_C\omega_C$. We illustrate the bound with an
example of state-independent contextuality and one of state-dependent
contextuality. In both, the bound holds with a wide margin.

\emph{Peres--Mermin square.} The Peres--Mermin
square~\cite{Peres90,Mermin90PM} is the standard example of
state-independent contextuality. Arrange nine two-qubit Pauli observables in
a $3\times3$ array so that the observables of each row and of each column
commute, the products along the rows and the first two columns are $+\one$,
and the product along the third column is $-\one$. The six rows and columns
are the contexts. With $\langle\cdot\rangle$ denoting the mean of the product
of the three outcomes of a row or column, the inequality $\beta=\sum_{\rm
rows}\langle\cdot\rangle+\sum_{\rm col\,1,2}\langle\cdot\rangle -\langle{\rm
col\,3}\rangle\le4$ holds for every assignment of values $\pm1$, whereas
$\beta(\rho)=6$ for every $\rho$. Each coefficient function takes the values
$\pm1$, so $\omega_C=2$. Theorem~\ref{thm:context} therefore gives
$\sum_C\TV\ge1$ for every schedule. For the maximally mixed state the minimum over all $9!$ schedules is $1.5$.
All two- and three-measurement records of these observables on $\one/4$ are
order independent, so the order dependence guaranteed by the theorem appears
only in schedules of at least four measurements. For example, with $XI=\sigma_x\otimes\one$, $IY=\one\otimes\sigma_y$, and so
on, the records of the schedules $(XI,IX,XX,IY)$ and $(XI,IX,IY,XX)$ are at
total-variation distance $1/2$.

The converse fails: order effects of this kind also occur without
contextuality. For three qubit observables $A$, $B$, $C$ along $\hat z$, $\hat y$, and $(\sqrt{1-t^2},0,t)$, the records
of $ABC$ and $BAC$ on $\one/2$ are at distance $t/2$ by
Theorem~\ref{thm:three}(a), although sharp qubit measurements admit no Kochen--Specker
contextuality.

\emph{KCBS pentagon.} The KCBS inequality~\cite{KCBS08} is the standard
state-dependent test of contextuality for a qutrit. For five rank-one
projectors $P_i$ on $\mathbb C^3$ with $P_iP_{i+1}=0$ (indices modulo $5$),
the contexts are the five pairs $\{i,i+1\}$, and $x_i\in\{0,1\}$ is the
outcome of the measurement $\{P_i,\one-P_i\}$. Orthogonal neighbors are
exclusive: the outcomes $x_i=x_{i+1}=1$ never occur together in a joint
measurement. The usual bound $\sum_i\langle P_i\rangle\le2$ holds only for
assignments that respect this exclusivity. A sequential distribution need not
respect it, because measurements performed between $P_i$ and $P_{i+1}$ can
disturb the state. We therefore use
$\beta'=\sum_i[p(x_i=1)-p(x_i=x_{i+1}=1)]\le2$, which holds for every 0/1
assignment and equals $\sum_i\langle P_i\rangle$ on quantum behaviors. Here
$\alpha_C(x_i,x_{i+1})=x_i-x_ix_{i+1}$ and $\omega_C=1$. At the KCBS optimum, where $\beta'=\sqrt5$, the smallest value of $\sum_C\TV$ over all $120$
schedules is $1.067$, whereas the theorem requires only
$\sqrt5-2\approx0.236$. Theorem~\ref{thm:kcbs} gives a bound in terms of
pairwise order scores.

The same rewriting works for every inequality
$\sum_iw_ip(x_i=1)\le\alpha(\mathcal G,w)$ with weights $w_i\ge0$ on an exclusivity graph $\mathcal G$, whose vertices
are the projectors and whose edges join mutually exclusive ones. The number $\alpha(\mathcal G,w)$, not to be confused with the coefficients
$\alpha_C$, is the weighted independence number, the largest total weight of a set of pairwise nonadjacent vertices. Subtracting $\sum_{ij\in E(\mathcal
G)}c_{ij}\,p(x_i=x_j=1)$ with $c_{ij}\ge\max(w_i,w_j)$ gives an inequality
that holds for every 0/1 assignment (Appendix~\ref{app:sequential}) and
coincides with the original one on quantum behaviors. The constants $\omega_C$ then depend on how the vertex and
edge terms are assigned to contexts.

\subsection{Steering and Bell nonlocality}\label{sec:spatial}
In a steering or Bell test each party measures one setting per round. A
classical model is obtained if a party instead measures all its settings in
a fixed order $\pi$ in every round and reports the recorded outcome of the
setting that is asked for. The report is exact for the setting measured
first. For another setting $x$ the error is the disturbance of $x$ by the settings
that precede it. It is bounded by the distance between the records of $\pi$
and of the schedule with $x$ moved to the front. The local state of the party depends on the outcomes of the other
parties, so we maximize this distance over local states.

For a party with measurements $\{\mathcal M_x\}$, where $x$ now labels
settings, and two local schedules $\pi,\pi'$ let
$\cT(\pi,\pi')=\max_\sigma\TV(p^\sigma_\pi,p^\sigma_{\pi'})$ over local states
$\sigma$, and let $\pi^x$ denote $\pi$ with setting $x$ moved to the front.
Unlike the distances in Theorem~\ref{thm:context}, these scores are properties of the local measurements alone and do not depend
on the shared state. We write $\cT_A(\pi,\pi')$ and $\cT_k(\pi,\pi')$ for this
quantity when the measurements are those of Alice and of party $k$.

\begin{theorem}[Steering and Bell nonlocality require local order effects]
\label{thm:spatial}
Let a multipartite state be measured locally with L\"uders instruments.
(a) \emph{Steering.} Let $\sigma_{a|x}=\Tr_A[(E_{a|x}\otimes\one)\rho]$ be Bob's assemblage, that
is, his unnormalized conditional states when Alice obtains outcome $a$ for
setting $x$. An assemblage has a local-hidden-state model if it has the form
$\tau_{a|x}=\sum_\lambda p(a|x,\lambda)\sigma_\lambda$ with positive operators
$\sigma_\lambda$ and conditional probabilities $p(a|x,\lambda)$, and steering
means that no such model exists. If Alice measures
all her settings in the schedule $\pi$, Bob's conditional states define a local-hidden-state assemblage
$\tau_{a|x}$ with
$\sum_a\|\sigma_{a|x}-\tau_{a|x}\|_1\le n_x\,\cT_A(\pi,\pi^x)$, where $n_x$ is
the number of outcomes of setting $x$. Hence every linear steering inequality
$\sum_{a,x}\Tr(\Phi_{a|x}\sigma_{a|x})\le b_{\rm LHS}$ with Hermitian witness
operators $\Phi_{a|x}$ is violated by at most
$\sum_xf_xn_x\cT_A(\pi,\pi^x)$, where
$f_x=\min_{\mu\in\mathbb R}\max_a\|\Phi_{a|x}-\mu\one\|$.
(b) \emph{Bell nonlocality.} Local schedules $\pi_k$ of all parties define a
local model whose distribution for the settings $\bm x$ differs from the
quantum one by at most $\sum_k\cT_k(\pi_k,\pi_k^{x_k})$ in total variation.
Scheduling all parties except party $j$ already gives a local model with error at
most $\sum_{k\ne j}\cT_k(\pi_k,\pi_k^{x_k})$. Hence a Bell inequality
$\sum_{\bm x}\sum_{\bm a}\alpha_{\bm x}(\bm a)p(\bm a|\bm x)\le b$ is violated
by at most $\min_j\sum_{\bm x}\omega_{\bm x}\sum_{k\ne
j}\cT_k(\pi_k,\pi_k^{x_k})$, where $\omega_{\bm x}=\max\alpha_{\bm
x}-\min\alpha_{\bm x}$. For CHSH, $\beta-2\le4\min\{\cT_A,\cT_B\}$ with $\cT_A=\cT(A_1,A_2)$ and
$\cT_B=\cT(B_1,B_2)$.
\end{theorem}

The constant $f_x$ plays the role of $\omega_C/2$ in Theorem~\ref{thm:context}. It is the
largest operator norm of the witness operators of setting $x$ after the
optimal shift by a multiple of the identity, which does not change the
violation because $\sum_a\sigma_{a|x}=\sum_a\tau_{a|x}$.
Theorem~\ref{thm:spatial} is a quantitative form, for sequential
implementations, of the general principle that jointly measurable settings
admit local-hidden-state and local-hidden-variable
models~\cite{Wolf09,Quintino14,Uola14}. Quantitative
versions based on approximate joint measurements and on incompatibility
robustness are given in Refs.~\cite{Andersson05,Cavalcanti16,Designolle19}.
The bounds of Theorem~\ref{thm:spatial} are linear in the order scores and
loose. For CHSH with an ideal partner, that is, with Bob's two settings chosen
optimally, the exact excess is
$2\sqrt{1+2\cT_A}-2\le2\cT_A$, at most half the bound of part (b).

\section{Product structure of Bell operators}\label{sec:product}
The bounds of Sec.~\ref{sec:spatial} are linear in the order scores of one
party and do not use the form of the Bell inequality. For correlation
inequalities with binary sharp observables a sharper statement is available.
Landau~\cite{Landau87} showed that the square of the CHSH operator is
$4\,\one$ minus the tensor product of the two local commutators, so that the
violation is controlled by the product of the two commutator norms. This
section extends the argument to all bipartite correlation functionals.

\subsection{The squared Bell operator}
Consider a correlation Bell functional for binary sharp observables,
$\beta=\sum_{x=1}^n\sum_{y=1}^mM_{xy}\langle A_x\otimes B_y\rangle$, where $M$
is a real matrix of coefficients for the $n$ settings of Alice and the $m$
settings of Bob. The Bell operator is
$\mathfrak B=\sum_{xy}M_{xy}A_x\otimes B_y=\sum_xA_x\otimes R_x$ with
$R_x=\sum_yM_{xy}B_y$. For the given observables the largest value of
$|\beta|$ over all states is $\bQ=\|\mathfrak B\|$. The classical bound is
$\bC=\max_{\bm s,\bm t}\bm s^{\mathsf T}M\bm t$ over sign vectors
$\bm s\in\{\pm1\}^n$ and $\bm t\in\{\pm1\}^m$, which are the deterministic
local assignments. For CHSH, $M$ has the rows $(1,1)$ and $(1,-1)$ and
$\bC=2$. We write $\cT^A_{xx'}=\cT(A_x,A_{x'})$ and
$\cT^B_{yy'}=\cT(B_y,B_{y'})$ for the order scores of the local pairs.

A product of two involutions splits into a Hermitian and an anti-Hermitian
part, $A_xA_{x'}=A_x\circ A_{x'}+\frac12[A_x,A_{x'}]$, where
$X\circ Y=\frac12(XY+YX)$ is the Jordan product. The square of the Bell
operator splits accordingly. Its \emph{Jordan part} collects the Jordan
products,
\begin{equation}
 \Gamma=\sum_{x,x'}\sum_{y,y'}M_{xy}M_{x'y'}(A_x\circ A_{x'})\otimes
 (B_y\circ B_{y'}),
 \label{eq:Jordan}
\end{equation}
and the remainder contains only commutators [Eq.~\eqref{eq:Bsquare} below].
If the observables of one party commute, then $\mathfrak B^2=\Gamma$.
The Jordan part contains no explicit commutator factors, but it still
depends on the potentially noncommuting observables and is not, by itself,
a classical Bell operator. The question is whether it can exceed $\bC^2$. We call
the inequality $\lambda_{\max}(\Gamma)\le\bC^2$ the \emph{Jordan estimate}.

We say that Alice's observables admit an \emph{operator cut decomposition}
if there are operators $\Omega_{\bm s}\succeq0$, indexed by sign vectors
$\bm s\in\{\pm1\}^n$ modulo a global sign, with $\sum_{\bm s}\Omega_{\bm s}=\one$
and $A_x\circ A_{x'}=\sum_{\bm s}s_xs_{x'}\Omega_{\bm s}$ for all $x,x'$.
The condition generalizes commutativity. For commuting involutions the joint
spectral projectors $\Pi_{\bm s}$ of the sign vectors $\bm s$ give
$A_xA_{x'}=\sum_{\bm s}s_xs_{x'}\Pi_{\bm s}$. An operator cut decomposition
requires only that the Jordan products have this form, with a POVM
$\{\Omega_{\bm s}\}$ in place of the joint projectors. The vectors $\bm s$
and $-\bm s$ give the same products $s_xs_{x'}$ and are identified. The name
refers to the matrices $\bm s\bm s^{\mathsf T}$, which are the vertices of the
cut polytope. With such a decomposition,
$\Gamma=\sum_{\bm s}\Omega_{\bm s}\otimes(\sum_xs_xR_x)^2$ is an operator
mixture of the squares of the operators $\sum_xs_xR_x$ that Bob faces when
Alice answers with the fixed signs $\bm s$. Each operator $\sum_xs_xR_x$ has
norm at most $\bC$, so the Jordan estimate holds.

\begin{theorem}[Product structure]\label{thm:product}
(a) The Bell operator satisfies
\begin{equation}
 \mathfrak B^2=\Gamma-\sum_{x,x'}\sum_{y,y'}M_{xy}M_{x'y'}\,
 \frac{[A_x,A_{x'}]}{2i}\otimes\frac{[B_y,B_{y'}]}{2i},
 \label{eq:Bsquare}
\end{equation}
and therefore, with sums over ordered pairs,
\begin{equation}
 \bQ^2\le\lambda_{\max}(\Gamma)+4\sum_{x\ne x'}\sum_{y\ne y'}
 |M_{xy}M_{x'y'}|\,\cT^A_{xx'}\cT^B_{yy'} .
 \label{eq:productbound}
\end{equation}
(b) If Alice's observables admit an operator cut decomposition, then
$\lambda_{\max}(\Gamma)\le\max_{\bm s}\|\sum_xs_xR_x\|^2\le\bC^2$; the same holds
with the parties exchanged. A decomposition exists for every pair of involutions, with
$\Omega_\pm=(\one\pm A_1\circ A_2)/2$, and for every set of commuting
involutions. For three involutions it exists exactly when
$(s_1A_1+s_2A_2+s_3A_3)^2\succeq\one$ for all signs.
(c) If $n=2$ and $\theta_j\in[0,\pi]$ are the angles of all blocks of $(A_1,A_2)$, including one-dimensional ones,
then
\begin{equation}
 \begin{aligned}
 \bQ^2=\max_{j,\pm}\lambda_{\max}\big[&R_1^2+R_2^2+\cos\theta_j\{R_1,R_2\}\\
 &\pm\sin\theta_j\,i[R_1,R_2]\big]
 \end{aligned}
 \label{eq:twosetting}
\end{equation}
and $\bQ^2\le\bC^2+8\cT^A_{12}\sum_{y\ne y'}|M_{1y}M_{2y'}|\cT^B_{yy'}$, with
equality for CHSH.
\end{theorem}

For CHSH, $\Gamma=4\,\one$ and Eq.~\eqref{eq:Bsquare} reduces to Landau's
identity, $\mathfrak B^2=4\,\one-[A_1,A_2]\otimes[B_1,B_2]$. In general the identity separates
$\mathfrak B^2$ into a part that does not require noncommutativity and a part
in which every term contains one commutator of each party. The norms of the
local commutators are $4\cT^A_{xx'}$ and $4\cT^B_{yy'}$, so the excess of
$\bQ^2$ over $\lambda_{\max}(\Gamma)$ is bounded by products of one order
score of each party [Eq.~\eqref{eq:productbound}]. A commutator expansion of
the Mermin operators is given in Ref.~\cite{Cereceda01}, and the relation
between local commutativity and Bell violation is discussed in
Ref.~\cite{Seevinck07}.

Part (b) is proved in Appendix~\ref{app:product}. For three involutions its
condition says that the correlations $\langle A_x\circ A_{x'}\rangle$ of
sequential measurements (Sec.~\ref{sec:bell64}) satisfy the Leggett--Garg
inequalities~\cite{LeggettGarg85} on every input. These inequalities hold
for a system with definite values that is not disturbed by measurement. Part (c) gives the exact
value when Alice has two settings. By Jordan's lemma (Sec.~\ref{sec:binary}) the pair $(A_1,A_2)$ is a direct sum
of blocks with angles $\theta_j$, and
Eq.~\eqref{eq:twosetting} reduces $\bQ$ to an eigenvalue problem for Bob's
operators $R_1$ and $R_2$ alone. The identity and the upper bounds of the
theorem also hold for a single system if the $A_x$ commute with the $B_y$,
with tensor products replaced by products of commuting operators.

Whenever the Jordan estimate holds and $M\ne0$, so that $\bC>0$,
Eq.~\eqref{eq:productbound} gives
\begin{equation}
 \bQ-\bC\le\frac2{\bC}\sum_{x\ne x'}\sum_{y\ne y'}|M_{xy}M_{x'y'}|\,
  \cT^A_{xx'}\cT^B_{yy'} .
 \label{eq:bilinear}
\end{equation}
The right-hand side vanishes if the order scores of either party vanish,
whereas the bound of Theorem~\ref{thm:spatial}(b) is linear in one party's
scores. The difference matters for the inverse question: given an observed value
$\beta>\bC$, how large must the order scores of the observables be? Corollary~\ref{cor:DI}
answers this question with the bilinear bound.

\begin{corollary}[Order scores of projective realizations]\label{cor:DI}
Let Alice have two settings, and let an observed value $\beta>\bC$ of a
correlation functional be reproduced by binary sharp observables (in any
dimension) with L\"uders instruments. Then these observables satisfy
$\cT^A_{12}\max_{y\ne y'}\cT^B_{yy'}\ge(\beta^2-\bC^2)/(8\sum_{y\ne y'}
|M_{1y}M_{2y'}|)$. For CHSH, $\cT_A\cT_B\ge(\beta^2/4-1)/4$, so that both order scores equal
$1/2$ at the Tsirelson value $\beta=2\sqrt2$.
\end{corollary}

The CHSH case follows from Landau's identity~\cite{Landau87,Seevinck07}. At
the Tsirelson value Theorem~\ref{thm:spatial}(b) gives only
$\min\{\cT_A,\cT_B\}\ge(\sqrt2-1)/2\approx0.207$, but it holds for arbitrary
L\"uders instruments. The corollary does not apply to the physical
instruments of an untrusted device. A Naimark dilation, which replaces a
POVM by a projective measurement on a larger space, reproduces the
statistics but changes the instruments, and order scores belong to
instruments. For unsharp L\"uders measurements the bound fails. With Alice's effects
$(\one\pm\eta\sigma_z)/2$ and $(\one\pm\eta\sigma_x)/2$, an ideal qubit
partner, and a maximally entangled state, one has $\beta=2\sqrt2\,\eta$. For
$\eta=0.9$ the corollary would require $\cT_A\cT_B\ge0.155$, whereas Eq.~\eqref{eq:unsharp}
gives $\cT_A\cT_B=0.127$.

\subsection{The general case}
For three or more settings for both parties the Jordan estimate can fail. On $\mathbb C^3\otimes\mathbb C^3$
let $A_1=B_1=\mathrm{diag}(-1,1,-1)$,
$A_2=B_2=\mathrm{diag}(-1,-1,1)$, $A_3=\one-2\bm v\bm v^{\mathsf T}$, and
$B_3=-A_3$, with $\bm v=(\sqrt p,\sqrt{(1-p)/2},\sqrt{(1-p)/2})$. As the functional take
the correlation part of the inequality $I_{3322}$ for three binary settings
per party~\cite{CollinsGisin04}, that is, its coefficient matrix without the
marginal terms. This is the matrix $M$ with rows $(1,1,1)$, $(1,1,-1)$,
$(1,-1,0)$, for which $\bC=4$. Then
$\lambda_{\max}(\Gamma)/\bC^2$ reaches $1.0346$ at $p\approx0.570$ and equals $1.0336$ at $p=1/2$. Since $\bQ=\bC$ throughout, the bilinear bound
itself holds trivially. We therefore state the
bound, not the Jordan estimate, as a conjecture.

\begin{conjecture}[Bilinear bound]\label{conj:product}
For every correlation functional and all binary sharp observables,
$\bQ^2\le\bC^2+4\sum_{x\ne x'}\sum_{y\ne y'}|M_{xy}M_{x'y'}|\,\cT^A_{xx'}
\cT^B_{yy'}$.
\end{conjecture}

Theorem~\ref{thm:product}(b) proves the conjecture when either party's
observables admit an operator cut decomposition. It also holds for two qubits
when one party has at most three settings (Proposition~\ref{prop:HSG}). The numerical evidence
beyond these cases is limited. We ran multistart local optimization of $\|\mathfrak B\|^2$ minus the right-hand
side of Conjecture~\ref{conj:product} over involutions whose $+1$ eigenspace is one dimensional (for $d=3$ these are
not all involutions), with four restarts for each $(d_A,d_B)\in\{(3,3),(2,3)\}$. The functionals
were the chained~\cite{BraunsteinCaves90} and $I_{3322}$ correlation
functionals and seven random $3\times3$ ones. The objective is zero at commuting configurations that attain $\bC$, and no
run gave a positive value. Among $4000$ random
instances with two or three settings per party and $d_A,d_B\le3$, $20$ of the $777$ instances outside the proved cases had $\bQ>\bC$, and their largest ratio of
$\bQ^2-\bC^2$ to the bilinear term was $0.89$.

On Clifford realizations and on product inputs the Jordan estimate reduces to
a problem about Gram matrices. Let $\gamma_1,\ldots,\gamma_r$ be pairwise
anticommuting Hermitian involutions, which exist on $\mathbb C^{2^{\lfloor
r/2\rfloor}}$. For every unit vector $\bm u_x\in\mathbb R^r$,
$A_x=\sum_iu_{x,i}\gamma_i$ is an involution, and $A_x\circ A_{x'}=(\bm
u_x\cdot\bm u_{x'})\one$. We call such observables, with unit vectors $\bm v_y$ for Bob, a
\emph{Clifford realization}. The $\gamma_i$ generate a Clifford algebra. For
$r=3$ they can be taken to be the Pauli matrices, and the observables are the
qubit observables $\bm u_x\cdot\bm\sigma$. Clifford realizations are a
standard test class because, by Tsirelson's theorem, the largest quantum
value of every correlation functional is attained on
them~\cite{Tsirelson80,Tsirelson87}. On a Clifford realization all Jordan
products are multiples of the identity, so $\Gamma$ is a multiple of the identity. Explicitly,
$\Gamma=\|\sum_{xy}M_{xy}\bm u_x\bm v_y^{\mathsf T}\|_F^2\,\one$, where
$\|\cdot\|_F$ is the Hilbert--Schmidt (Frobenius) norm. For arbitrary involutions and a
product input $\ket\alpha\otimes\ket\beta$,
$\langle\Gamma\rangle=\Tr(G_AMG_BM^{\mathsf T})$, where $G_A$ and $G_B$ are
the real parts of the Gram matrices of the unit vectors $A_x\ket\alpha$ and
$B_y\ket\beta$.

The Jordan estimate on Clifford realizations is therefore the
inequality~\eqref{eq:HSG} below. It resembles Grothendieck's inequality,
which states that $\sum_{xy}M_{xy}\bm u_x\cdot\bm v_y\le K_G\,\bC$ for all
unit vectors, with a universal constant $K_G>1$~\cite{Krivine79,Briet14}. In
Bell language, $K_G$ bounds the ratio of the quantum to the classical value
of correlation functionals~\cite{Tsirelson87}, and $K_G(r)$ is the best
constant for vectors in $\mathbb R^r$. The left-hand side of Grothendieck's
inequality is the trace of the matrix
$\sum_{xy}M_{xy}\bm u_x\bm v_y^{\mathsf T}$. Inequality~\eqref{eq:HSG} replaces
the trace by the Hilbert--Schmidt norm and asserts the constant $1$. We call
a real matrix $Z$ a local correlation matrix if it lies in
$\mathrm{conv}\{\bm s\bm t^{\mathsf T}\}$, the set of correlation matrices of
local-hidden-variable models.

\begin{proposition}[Hilbert--Schmidt reduction and proved cases]\label{prop:HSG}
For real $n\times m$ matrices $M$ consider the inequality
\begin{equation}
 \Big\|\sum_{xy}M_{xy}\bm u_x\bm v_y^{\mathsf T}\Big\|_F\le\bC
 \quad\text{for all unit vectors }\bm u_x,\bm v_y .
 \label{eq:HSG}
\end{equation}
(i) Its validity for all $M$ is equivalent to each of the following, also
for all $M$: $\Tr(G_AMG_BM^{\mathsf T})\le\bC^2$ for all real Gram matrices
$G_A,G_B$ of unit vectors; the matrix $Z_{xy}=\bm u_x^{\mathsf T}W\bm v_y$ is a
local correlation matrix whenever $\|W\|_F\le1$; and
$\langle\Gamma\rangle\le\bC^2$ on all product inputs.
(ii) It holds with $\bC$ replaced by $\sqrt{\pi/2}\,\bC$.
(iii) It holds for all $M$ whenever $\min(n,m)\le3$.
(iv) It implies $K_G(r)\le\sqrt r$ for the real Grothendieck constant of order
$r$, in particular Krivine's bound $K_G(2)\le\sqrt2$~\cite{Krivine79}.
(v) For vectors in $\mathbb R^2$ and $n,m\ge2$, it is equivalent to the
following statement, proved by part (iii) if $\min(n,m)\le3$ and not
established here otherwise. A two-qubit state with vanishing local Bloch
vectors and correlation tensor $\mathrm{diag}(c,s)$ in a fixed measurement
plane gives local correlations
$c\cos\alpha_x\cos\beta_y+s\sin\alpha_x\sin\beta_y$ for \emph{every choice}
of $n$ and $m$ directions in that plane if and only if $c^2+s^2\le1$.
The latter condition is equivalent to absence of CHSH violation optimized
over all directions in that plane~\cite{Horodecki95}; it is not a criterion
for locality of an arbitrary fixed finite set of directions.
\end{proposition}

Part (iii) follows from the facial structure of the correlation polytope
$\mathrm{conv}\{\bm s\bm t^{\mathsf T}\}$: for $\min(n,m)\le3$ its facets are the trivial inequalities $|Z_{xy}|\le1$ and the CHSH
inequalities, by a
theorem of Barahona and Mahjoub~\cite{Barahona86}, and the matrices $Z$ of
part (i) satisfy both. By part
(iii) and
Theorem~\ref{thm:product}(a), Conjecture~\ref{conj:product} holds on Clifford realizations whenever one
party has at most three settings. For $r=3$ these are the two-qubit observables $\bm u_x\cdot\bm\sigma$ and $\bm
v_y\cdot\bm\sigma$. By part (ii) the conjecture
holds on all Clifford realizations with $\bC^2$ replaced by $(\pi/2)\bC^2$. For $n,m\ge4$ the correlation polytope has other facets, and
Eq.~\eqref{eq:HSG} is open.

The isotropic case $W\propto\one$ follows from
known bounds on $K_G(r)$~\cite{Krivine79,Hirsch17,Designolle23}, and the planar isotropic case is the local model of Ac\'in, Gisin, and
Toner~\cite{AcinGisinToner06}. In the $(c,s)$ plane of part (v), locality is elementary for $|c|+|s|\le1$,
where $Z$ is a mixture of product correlations, and it holds at the four
isotropic points $|c|=|s|=1/\sqrt2$ by the model just mentioned. The convex
hull of these sets is a regular octagon inscribed in the unit disk, whereas
Eq.~\eqref{eq:HSG} asserts locality on the whole disk.

Floating-point linear programs over all deterministic strategies support locality in part (v)
for grids of eight equally spaced directions for each party and all tested
$(c,s)$ with $c^2+s^2=1$. For four settings per party, a Nelder--Mead search
over all angles and over $c^2+s^2=1$ found no $Z$ outside $(1+10^{-8})\,\mathrm{conv}\{\bm s\bm t^{\mathsf T}\}$,
as estimated by those linear programs. Solver status and primal and dual
residuals are checked in the revised code. These are finite-precision
checks, not exact-arithmetic certificates or a proof for all angles.

\subsection{Steering with an anticommuting partner}
In a steering test Bob's measurements are trusted and known, and Alice's are
uncharacterized. We consider the linear steering functional
$\langle\sum_xA_x\otimes B_x\rangle$~\cite{Cavalcanti09} with pairwise
anticommuting observables of Bob, for which every pair has the maximal order
score $1/2$. Bob's commutators are then fixed, so the bound that corresponds
to Eq.~\eqref{eq:productbound} is linear in the order scores of Alice.

\begin{theorem}[Steering bound]\label{thm:steer}
Let Bob's trusted observables $B_1,\ldots,B_n$ be pairwise anticommuting
Hermitian involutions (for example, Pauli observables for $n\le3$). Then for all binary
sharp $A_x$,
\begin{equation}
 \max_\rho\Big|\Big\langle\sum_xA_x\otimes B_x\Big\rangle\Big|
 \le\Big(n+4\sum_{x<x'}\cT^A_{xx'}\Big)^{1/2},
 \label{eq:steern}
\end{equation}
with equality for $n=2$ and all $A_1$, $A_2$, and for $n=3$ when Alice also
measures three anticommuting Pauli observables. Local-hidden-state models
satisfy $\langle\sum_xA_x\otimes B_x\rangle\le\sqrt n$~\cite{Cavalcanti09}.
\end{theorem}

The local-hidden-state bound $\sqrt n$ is the value of the right-hand side
of Eq.~\eqref{eq:steern} when all order scores of Alice vanish. A violation of the local-hidden-state bound therefore requires order
dependence between at least one pair of Alice's observables, and each pair
adds at most $4\cT^A_{xx'}\le2$ to the bound on the squared value.

\subsection{Minimal scenarios}
With two binary sharp settings per party the bounds of Theorem~\ref{thm:steer}
and, for CHSH, of Theorem~\ref{thm:product}(c) become equalities, and the
maximal steering, CHSH, and MABK values are functions of the order scores
alone. Figure~\ref{fig:minimal}(a) shows the
first two.

\begin{corollary}[Two binary settings per party]\label{cor:minimal}
Let $\cT_X=\|[X_1,X_2]\|/4$ for party $X$.
(i) $\max_\rho\langle A_1\otimes\sigma_x+A_2\otimes\sigma_z\rangle=
\sqrt{2+4\cT_A}$.
(ii) $\max_\rho|\langle\mathfrak B_{\rm CHSH}\rangle|
=\|\mathfrak B_{\rm CHSH}\|=2\sqrt{1+4\cT_A\cT_B}$
\cite{Landau87,Kiukas10}, which becomes $2\sqrt{1+2\cT_A}$ when Bob's
qubit settings are optimized. For $N$ parties with two settings $X_{k,1},X_{k,2}$ for party $k$, the
Mermin--Ardehali--Belinskii--Klyshko (MABK) operator is the standard
generalization of the CHSH operator~\cite{GisinBechmann98,WernerWolf01}. It
is defined by $\mathfrak M_1=X_{1,1}$ and
$\mathfrak M_N=\frac12[\mathfrak M_{N-1}\otimes(X_{N,1}+X_{N,2})+
\mathfrak M'_{N-1}\otimes(X_{N,1}-X_{N,2})]$, where the prime exchanges
$X_{k,1}\leftrightarrow X_{k,2}$ for all parties. It satisfies
$\|\mathfrak M_N\|^2=\frac12[\prod_k(1+2\cT_k)+\prod_k(1-2\cT_k)]$; the local
bound is $1$.
(iii) For two qubits with concurrence $C$~\cite{Wootters98}, an entanglement
measure that ranges from $0$ for separable to $1$ for maximally entangled
states, the CHSH value
$\beta$ satisfies $\beta^2/4-1\le\min\{C^2,4\cT_A\cT_B\}$.
Each bound is separately tight; simultaneous saturation for arbitrary
prescribed $C,\cT_A,\cT_B$ is not asserted.
\end{corollary}

Part (i) is the case $n=2$ of Theorem~\ref{thm:steer}, and part (ii) for CHSH
is the equality case of Theorem~\ref{thm:product}(c). The MABK formula follows
directly from the above recursion, as shown in Appendix~\ref{app:product}.
The related Mermin-operator expansion in Ref.~\cite{Cereceda01} uses a
different operator for even particle number. For $\cT_k=1/2$ for all parties it gives the maximal
value $\|\mathfrak M_N\|=2^{(N-1)/2}$. Part (iii) combines part (ii) with the
Verstraete--Wolf bound~\cite{Verstraete02}. It displays two resources, the
entanglement of the state through $C$ and the order dependence of the
measurements through $\cT_A\cT_B$, and the CHSH violation is limited by the
smaller one. Device-independent state-estimation methods based on CHSH are
discussed in Ref.~\cite{Bardyn09}, but neither the two-qubit concurrence in
part (iii) nor the order scores of the physical instruments follow without
the stated dimensional and instrument assumptions.

\begin{table*}[t]
\caption{Quantitative necessary conditions obtained in this article (notation of
Secs.~\ref{sec:uncert}--\ref{sec:three}). $\cT,\cJ$: order scores of a pair (maximized over
inputs). Schedule TV: total-variation distance between the records of two
schedules on the input that shows the signature. $\cT(\pi,\pi')$: the same
distance maximized over local states. Rows 2 and 4--7 assume binary sharp observables, except for the bounds with
$\cT(\pi,\pi')$. The entry marked $^{*}$ is proved when one party's observables
admit an operator cut decomposition (e.g., two settings) and conjectured in
general.}
\label{tab:summary}
\begin{ruledtabular}
\begin{tabular}{p{0.16\textwidth}p{0.14\textwidth}p{0.41\textwidth}p{0.19\textwidth}}
Signature & Order quantity & Bound & Exact case\\
\hline
\raggedright Entropic uncertainty & $\cT$, $\cJ$ & \raggedright
 ${\cT\ge1-2^{-\qmu}}$, ${\cJ\ge\varphi(2^{-\qmu})}$,
 ${\cJ\ge\varphi(1-g_k^{-1}(\Ustar))}$ & MUBs ($\cT$); qubit MUBs ($\cJ$)\\
\raggedright Variance, Robertson & $\cT$ & \raggedright
 ${U_{\rm var}\le1-\sqrt{1-4\cT^2}}$;
 ${\max_\rho|\langle[A,B]\rangle|/2=2\cT}$ & qubit pairs\\
Contextuality & schedule TV & \raggedright
 ${\beta-b\le\sum_C\omega_C\TV(p^\rho_\pi|_C,p^\rho_{\pi_C}|_C)}$ & $\cdots$\\
Steering & \raggedright $\cT(\pi,\pi')$; $\cT^A_{xx'}$ & \raggedright
 violation ${\le\sum_xf_xn_x\cT_A(\pi,\pi^x)}$; Eq.~\eqref{eq:steern} &
 $n=2$: $\sqrt{2+4\cT_A}$\\
Bell nonlocality & \raggedright $\cT(\pi,\pi')$; $\cT^A,\cT^B$ & \raggedright
 two parties: violation ${\le\min_k\sum_{\bm x}\omega_{\bm x}\cT_k}$;
 ${\bQ^2-\bC^2\le4\sum|MM|\cT^A\cT^B}$\,$^{*}$ &
 CHSH: $2\sqrt{1+4\cT_A\cT_B}$\\
\raggedright Bell 1964 on $\ket\Phi$ & \raggedright records of $\one/d$; $\cT$ &
 \raggedright ${\beta_3-1=\kappa_B-4p_{(B)}(a\ne b\ne c)}$,
 ${\kappa_B\le4\cT(A,B)\cT(B,C)}$ & identity\\
\raggedright Rank-one KCBS on $\mathbb C^3$ & \raggedright $\cT_i$ of chords & \raggedright
 ${V\le\max_i\cT_i\cT_{i+1}}$ & KCBS optimum\\
\end{tabular}
\end{ruledtabular}
\end{table*}

\section{Three measurements and the KCBS pentagon}\label{sec:three}
The bounds of Sec.~\ref{sec:product} involve two incompatible pairs held by
different parties. Two further scenarios obey bounds of the same form, with
both pairs taken from one set of observables: three binary sharp measurements with Bell's original inequality~\cite{Bell64},
and the KCBS pentagon~\cite{KCBS08}.

\subsection{Three binary sharp measurements}\label{sec:bell64}
Bell's original inequality~\cite{Bell64} bounds a combination of three
correlators, $E_{AB}+E_{BC}-E_{AC}\le1$, where $E_{XY}$ is the correlation of
Alice's outcome for $X$ with Bob's outcome for $Y$. It holds for local models
with perfect correlations, which a maximally entangled state provides. The
local state of a maximally entangled state is maximally mixed, and on this
input the records of two measurements do not depend on the order
[Proposition~\ref{prop:basic}(iv)]. The simplest order effect on this input therefore
involves three measurements. We show that it governs the violation.

For Hermitian involutions $X$ and $Y$ write $XY=X\circ Y+iK_{XY}$ with
$K_{XY}=[X,Y]/(2i)$. The operators $X\circ Y$ and $iK_{XY}$ are the Hermitian
and anti-Hermitian parts of the product, as in Sec.~\ref{sec:product}, and
$\|K_{XY}\|=2\cT(X,Y)$ by Theorem~\ref{thm:binary}. The correlation of the two outcomes is
$\langle X\circ Y\rangle$ in either order~\cite{Fritz10}, so the order affects only the marginals. For three involutions the relation $AC=(AB)(BC)$ gives
\begin{equation}
 \Tr(AC)=\Tr[(A\circ B)(B\circ C)]-\Tr(K_{AB}K_{BC}).
  \label{eq:compose}
\end{equation}
Divided by $d$, the left-hand side is the correlation of $A$ and $C$ measured
in direct succession on $\one/d$, and the first term on the right is their
correlation when $B$ is measured between them. The two differ by a product
of the two commutators. For a schedule $(X,Y,Z)$ of the three measurements on
$\mathbb C^d$ we write $\kappa_Y=\Tr(K_{XY}K_{YZ})/d$, which is symmetric in
the outer measurements $X$ and $Z$. It is the change of the correlation of
the outer outcomes $x$ and $z$ caused by the middle measurement: in the
record of $(X,Y,Z)$ on $\one/d$, $\mathbb E[xz]=\Tr(XZ)/d+\kappa_Y$. It
vanishes if $Y$ commutes with $X$ or with $Z$.

\begin{theorem}[Three binary sharp measurements]\label{thm:three}
Let $A,B,C$ be Hermitian involutions on $\mathbb C^d$ with L\"uders
instruments.
(a) For $\rho=\one/d$ the record $p_{(Y)}$ of a schedule depends only on its
middle measurement $Y$, and for $Y\ne Y'$
\begin{equation}
 \TV(p_{(Y)},p_{(Y')})=\tfrac12\max\{|\kappa_Y|,|\kappa_{Y'}|\},
 \label{eq:tv3}
\end{equation}
where $|\kappa_B|\le4\,\cT(A,B)\,\cT(B,C)$.
(b) Let $\ket\Phi=d^{-1/2}\sum_j\ket{jj}$, and let
$E_{XY}=\langle\Phi|X\otimes Y^{\mathsf T}|\Phi\rangle=\Tr(XY)/d$ be the
correlation when Alice measures $X$ and Bob measures $Y^{\mathsf T}$, the
transpose of $Y$ in the basis $\{\ket j\}$. The
transpose gives perfect correlations, $E_{XX}=1$. Then
\begin{equation}
 E_{AB}+E_{BC}-E_{AC}=1+\kappa_B-4\,p_{(B)}(a\ne b\ne c),
 \label{eq:bell64}
\end{equation}
where $a\ne b\ne c$ stands for the two alternating outcome strings $(+,-,+)$
and $(-,+,-)$.
\end{theorem}

We write $\beta_3$ for the left-hand side of Eq.~\eqref{eq:bell64}, so that
Bell's original inequality reads $\beta_3\le1$. Since $E_{XY}=\Tr(XY)/d$, it
is also the Leggett--Garg combination of the sequential correlations of $A$,
$B$, and $C$ on the maximally mixed state, which is bounded by $1$ for a
system with definite values that is not disturbed by
measurement~\cite{LeggettGarg85,Fritz10,Budroni13}. By the relation $\mathbb E[xz]=\Tr(XZ)/d+\kappa_Y$, $\kappa_B$ is the mean of
$ac$ in $p_{(B)}$ minus its mean in $p_{(C)}$. It is a signaling-in-time term~\cite{KoflerBrukner13}, that is, a dependence
of later statistics on whether an earlier measurement was performed. The last term of Eq.~\eqref{eq:bell64} is never positive. With part (a) this
gives
\begin{equation}
 \beta_3-1\le\kappa_B\le4\,\cT(A,B)\,\cT(B,C),
 \label{eq:bell64bound}
\end{equation}
and also $\kappa_B\le2\,\TV(p_{(B)},p_{(C)})$. Equation~\eqref{eq:bell64bound}
states that a violation of Bell's original inequality by $\ket\Phi$ requires
both pairs $(A,B)$ and $(B,C)$ to be order dependent, and that its size is at
most four times the product of their order scores. The pair $(A,C)$ does not
enter.

In the Bell experiment, a measurement of $A^{\mathsf T}$ on one half of
$\ket\Phi$ with outcome $a$ leaves the other half in $\rho_a=P_a/\Tr P_a$ with
$P_a=(\one+aA)/2$. Conditioned on $a$, the records $p_{(B)}$ and $p_{(C)}$ are therefore the two
order records of the pair $(B,C)$ on the steered state $\rho_a$, and
$\TV(p_{(B)},p_{(C)})=\sum_ap(a)\,\cT(B,C|\rho_a)$ is the average order effect of the pair $(B,C)$.

For every state with $E_{BB}=1$, the quantity $1+\beta_3$ is a CHSH value for the settings
$(A,B)$ of Alice and $(B^{\mathsf T},C^{\mathsf T})$ of Bob, so
Corollary~\ref{cor:minimal}(ii) gives
$\beta_3\le2[1+4\cT(A,B)\cT(B,C)]^{1/2}-1$. This implies the outer inequality
of Eq.~\eqref{eq:bell64bound}, and Eq.~\eqref{eq:bell64} adds the exact
decomposition for $\ket\Phi$. For qubits, $\beta_3$ reaches its maximum $3/2$
at $\cT(A,B)=\cT(B,C)=\sqrt3/4$. Anticommuting pairs give
$\beta_3=-E_{AC}\le1$ in every dimension, so maximal order scores give no
violation.

The maximally mixed input is essential for Eq.~\eqref{eq:bell64bound}. For
general inputs the
Leggett--Garg operator $A\circ B+B\circ C-A\circ C-\one$ contains a term of
first order in the commutators, and its expectation can reach $1/2$ for
commuting $A$ and $B$ (Appendix~\ref{app:three}). One then has only a schedule bound analogous to Theorem~\ref{thm:context}.
Since $p^\rho_{ABC}$ is
a joint distribution with the correct marginal on $(a,b)$,
$\langle A\circ B+B\circ C-A\circ C\rangle_\rho\le
1+2\TV(p^\rho_{ABC}|_{bc},p^\rho_{BC})+2\TV(p^\rho_{ABC}|_{ac},p^\rho_{AC})$.

\subsection{The KCBS pentagon}\label{sec:kcbs}
Three sharp measurements cannot violate a noncontextuality inequality,
because their compatibility graph, whose vertices are the measurements and
whose edges join commuting pairs, is complete or acyclic~\cite{Budroni22}. On
a qutrit, the smallest scenario with
state-dependent contextuality is the KCBS pentagon (Sec.~\ref{sec:context}). Let
$v_1,\ldots,v_5\in\mathbb C^3$ be unit vectors with
$\langle v_i,v_{i+1}\rangle=0$ (indices modulo $5$), let $P_i=\proj{v_i}$, and
let $V=\lambda_{\max}(\sum_iP_i)-2$ be the largest violation of
$\sum_i\langle P_i\rangle\le2$. Place the five projectors on the vertices of a pentagon whose edges join
neighbors. Neighbors are orthogonal and hence compatible. The remaining five
pairs $(i,i+2)$ are the diagonals of the pentagon, which we call \emph{chords}
and label by $i$. Only they can be incompatible. Chord $i$ has the overlap
$z_i=|\langle v_i,v_{i+2}\rangle|^2$ and, by Theorem~\ref{thm:binary}, the
order score $\cT_i=\cT(P_i,P_{i+2})=[z_i(1-z_i)]^{1/2}$, where $P_i$ stands
for the binary measurement $\{P_i,\one-P_i\}$. Chords $i$ and $i+1$ cross,
whereas chords $i$ and $i+2$ share a vertex. The following theorem expresses
the violation $V$ through the chord overlaps [parts (a) and (b)] and bounds it
by the order scores of two crossing chords [part (c)].

\begin{theorem}[KCBS pentagon on $\mathbb C^3$]\label{thm:kcbs}
(a) $z_{i+3}=(1-z_i)(1-z_{i+1})$ for all $i$.
(b) $\sum_iz_i\le2$, and $V>0$ if and only if the inequality is strict; in
every case
\begin{equation}
 2-\sum_iz_i=\frac{V^2(1+V)}{1-V}.
 \label{eq:kcbsvalue}
\end{equation}
(c) For all $i$, $\cT_i\cT_{i+1}=V\,[(1+V)(1-z_{i+3})/(1-V)]^{1/2}$, and
\begin{equation}
 V\le\max_i\,\cT_i\,\cT_{i+1},
 \label{eq:kcbslaw}
\end{equation}
with equality for $V>0$ only if $z_i=(3-\sqrt5)/2$ for all $i$, the KCBS
optimum, where $V=\cT_i^2=\sqrt5-2$.
\end{theorem}

The relations (a) are those of the self-polar spherical pentagon, Gauss's
\emph{pentagramma mirificum}~\cite{Coxeter71}. Part (b) restates the characteristic polynomial of $\sum_iP_i$ found in
Ref.~\cite{Badziag11}. It says that the KCBS inequality is violated exactly
when the five chord overlaps sum to less than $2$. Part (c) is tight at the maximal violation. It is also tight asymptotically as
$V\to0$, because the formula for $\cT_i\cT_{i+1}$ gives
$\max_i\cT_i\cT_{i+1}\le V[(1+V)/(1-V)]^{1/2}$. Commutator-based
measures of contextuality in this scenario were considered in Ref.~\cite{Gunhan25}. With $A_i=\one-2P_i$, Eq.~\eqref{eq:kcbslaw} reads
$-\sum_i\langle A_iA_{i+1}\rangle-3\le4\max_i\cT_i\cT_{i+1}$, which has the form
of the CHSH bound $\beta-2\le2(1+4\cT_A\cT_B)^{1/2}-2\le4\cT_A\cT_B$ and of
Eq.~\eqref{eq:bell64bound}. In CHSH the two pairs are the diagonals of the square formed by Alice's and
Bob's observables in alternation, and in
the pentagon they are two crossing chords. In every dimension a violation requires two crossing noncommuting chords, as
the following proposition shows.

\begin{proposition}[Necessary chords]\label{prop:cross}
Let $A_1,\ldots,A_5$ be Hermitian involutions on $\mathbb C^d$ with
$[A_i,A_{i+1}]=0$, $P_i=(\one-A_i)/2$, and $S=\sum_iA_iA_{i+1}$, so that noncontextual models satisfy $-\langle S\rangle\le3$. We call the
chord $(i,i+2)$ noncommuting if $[A_i,A_{i+2}]\ne0$.
(a) If some $A_j$ commutes with all $A_i$, then $-S\preceq3\,\one$; hence a
violation requires at least three noncommuting chords, two of which cross.
(b) If $P_iP_{i+1}=0$ for all $i$ and some chord commutes, then
$\lambda_{\max}(\sum_iP_i)\le2$; hence a violation requires all five chords
to be noncommuting.
\end{proposition}

We conjecture that the bound~\eqref{eq:kcbslaw} extends to projectors of any
rank in every dimension.

\begin{conjecture}[Pentagon bound]\label{conj:kcbs}
For projectors $P_1,\ldots,P_5$ of arbitrary rank on $\mathbb C^d$ with
$P_iP_{i+1}=0$,
$\lambda_{\max}(\sum_iP_i)-2\le\max_i\cT(P_i,P_{i+2})\,\cT(P_{i+1},P_{i+3})$.
\end{conjecture}

Theorem~\ref{thm:kcbs} proves the conjecture for rank-one projectors on
$\mathbb C^3$ and for direct sums of such pentagons, because $V$ and the
$\cT_i$ of a direct sum are maxima over the blocks. By
Proposition~\ref{prop:cross}(b) the right-hand side is positive whenever the
left-hand side is. Among $4\times10^5$ random rank-one pentagons in arbitrary dimension, i.e.,
Gram matrices $\one+H\succeq0$ with $H$ supported on the chord cycle, the
largest ratio of the two sides was $0.90$. A random local search maximized this ratio at $V\ge0.02$ over pentagons with
orthogonal neighbors, with dimension and rank $(4,1)$, $(5,1)$, $(6,2)$,
$(7,2)$, $(9,3)$ and with ranks $(1,1,1,1,2)$ in $d=5$. It found ratios up to
$0.9998$ near embedded KCBS optima and none above $1$. This evidence is limited. The case of neighbors that commute without being
orthogonal and the case of longer odd cycles are open.

\section{Several measurements and unsharp measurements}\label{sec:more}
This section delimits what order scores detect, first for more than two
measurements and then for unsharp measurements.

\emph{Several measurements.} By Proposition~\ref{prop:basic}(iii), the order information $\cJ_\Sigma$ of
all $m!$ schedules vanishes for every input exactly when effects of different
measurements commute. For projective measurements this is pairwise
compatibility. The criterion does not capture genuinely multiobservable
incompatibility of POVMs, such as that of triples that are pairwise but not
triplewise jointly measurable~\cite{Kunjwal14}.

For sharp qubit observables $A_i=\bm a_i\cdot\bm\sigma$ with Gram matrix
$G_{ij}=\bm a_i\cdot\bm a_j$, L\"uders updates make the record a Markov chain in which consecutive outcomes
$s_i,s_j$ have correlation $\mathbb E[s_is_j]=G_{ij}$. Two-measurement records
therefore determine $G$, including its signs, and with it the tight variance
bound $\sum_i\Var(A_i)\ge m-\lambda_{\max}(G)$~\cite{Abbott16}. The
pairwise order \emph{scores} $\cT(A_i,A_j)=\frac12(1-G_{ij}^2)^{1/2}$ depend
only on $|G_{ij}|$ and do not determine three-observable uncertainty. For example, the $3\times3$ Gram matrices with all off-diagonal entries equal to $+1/3$
and to $-1/3$ have the same pairwise scores $\sqrt2/3$ but variance bounds $4/3$ and $5/3$. The pairwise records thus contain more
information than the pairwise scores.

\emph{Unsharp measurements.} Commutativity of the effects, the criterion of
Proposition~\ref{prop:basic}(iii)~\cite{Gudder01}, is strictly stronger than
joint measurability. For the unsharp qubit measurements with effects
$E_\pm=(\one\pm\eta\sigma_z)/2$ and $E'_\pm=(\one\pm\zeta\sigma_x)/2$, which
are noisy versions of $\sigma_z$ and $\sigma_x$ with sharpness parameters
$\eta,\zeta\in[0,1]$,
\begin{equation}
 \cT=\tfrac12\max\{\eta(1-\sqrt{1-\zeta^2}),\,\zeta(1-\sqrt{1-\eta^2})\},
 \label{eq:unsharp}
\end{equation}
which is positive for all $\eta,\zeta>0$, although the pair is jointly measurable for
$\eta^2+\zeta^2\le1$~\cite{Busch86,Yu10}. The L\"uders instrument of an
unsharp measurement still partially dephases the state, and this changes the
statistics of the other measurement. For POVMs a positive order score therefore does not imply incompatibility.
The steering and Bell bounds still give necessary order dependence under
their stated assumptions. The uncertainty statements require sharp
measurements: two commuting fair-coin POVMs with effects $\one/2,\one/2$
have entropy sum two bits on every input but zero order scores. Likewise,
the contextuality bounds concern the specified commuting-effect contexts,
not arbitrary joint implementations of compatible POVMs.

\section{Discussion}\label{sec:discussion}
Table~\ref{tab:summary} summarizes the necessary conditions obtained from
order records. Entropic, variance, and Robertson uncertainty relations for sharp measurements give lower bounds on
the order scores of a pair. A violated noncontextuality inequality requires
two schedules to differ on the same input, and steering and Bell violations
require local order effects. All these bounds have explicit constants, and
the corresponding converses need not hold for a fixed state and scenario. For correlation Bell inequalities with binary sharp
observables, the commutator part of the squared Bell operator is bounded
bilinearly in the two parties' order scores, and so is the violation whenever
one party's observables admit an operator cut decomposition. For Bell's original inequality on a maximally entangled state, the violation
is at most an order effect of three measurements, which obeys a bilinear bound. For the rank-one KCBS pentagon on a qutrit, a bound of the same form is
attained at the maximal violation.

The main open question is Conjecture~\ref{conj:product}, for which the route through the Jordan part
fails in general. On Clifford realizations and product inputs the Jordan
estimate is equivalent to the Hilbert--Schmidt Grothendieck
inequality~\eqref{eq:HSG}, from which the conjecture would follow on
Clifford realizations. Equation~\eqref{eq:HSG} is proved when one party has at most three settings
and is open otherwise. Its planar case would show that the correlations of
planar measurements on a two-qubit state with vanishing local Bloch vectors
are local for every choice of planar settings whenever the state satisfies
all planar CHSH inequalities, including unmeasured directions. Conjecture~\ref{conj:kcbs} for the pentagon is open, as are the case of
neighbors that commute without being orthogonal and the case of longer odd
cycles. The form of a bilinear bound for three measurements on general inputs,
where Eq.~\eqref{eq:bell64bound} fails, is also open.
Further open problems are the monotonicity of $\Psi$, equality on the left
of Eq.~\eqref{eq:mubJ} for MUBs with $d\ge3$, and the behavior of order
scores under coarse-graining of measurements. The order records of Sec.~\ref{sec:context} can be measured in the sequential
experiments used to test contextuality~\cite{Guhne10}.

\emph{Data availability.} The Supplemental Material~\cite{SM} contains the
code that reproduces all numerical statements and Fig.~\ref{fig:minimal}.

\begin{acknowledgments}
This work was supported by the National Natural Science Foundation of China
(Grant No.~12475020), the National Key Research and
Development Program of China (Grant No.~2023YFC2205802), and the Innovation
Program for Quantum Science and Technology (Grant No.~2021ZD0301701).
\end{acknowledgments}

\appendix
\section{Proofs for Sec.~\ref{sec:records}}\label{app:basic}
\emph{Proposition~\ref{prop:basic}.} (i) $\Djs$ is the mutual information
between a fair bit and the record, so it lies in $[0,1]$. For the lower bound
in Eq.~\eqref{eq:JT}, coarse-grain both records to the event $S$ that realizes
their total-variation distance; this gives Bernoulli distributions with
means $u$ and $u-\TV$. Mix this pair with the copy in which the two outcomes and the two records are
exchanged. By joint convexity and symmetry this does not increase $\Djs$, and
it produces Bernoulli means $(1\pm\TV)/2$, whose divergence is
$1-h((1+\TV)/2)$. For the upper bound
write $\Djs=\sum_i\frac{x_i+y_i}2[1-h(\frac{x_i}{x_i+y_i})]$ and use
$1-h(\frac{1+u}{2})\le|u|$. For Eq.~\eqref{eq:JH} compare the contributions of a
single cell $(x_i,y_i)=(y_iu^2,y_i)$. In nats they are $\frac{y_i}2f(u)$ for
$\Djs$, with $f(u)=u^2\ln w+\ln\frac2{1+u^2}$ and $w=\frac{2u^2}{1+u^2}$, and
$\frac{y_i}2(u-1)^2$ for $\cH^2$; cells with $y_i=0$ contribute
$\frac{x_i}2\ln2$ and $\frac{x_i}2$. One finds $f'(u)=2u\ln w$ and
$f''(u)=2\ln w+4/(1+u^2)$. The function $f_+=(u-1)^2-f$ satisfies
$f_+''=2[(w-1)-\ln w]\ge0$ and $f_+(1)=f_+'(1)=0$, hence $f_+\ge0$. The function
$f_-=f-\ln2\,(u-1)^2$ satisfies $f_-''=2[\ln(1-v)+2v]$ with $v=1/(1+u^2)$, so $f_-$ is
concave on $[0,u_0]$ and convex on $[u_0,\infty)$, $u_0\approx0.505$. Since
$f_-(0)=f_-(1)=f_-'(1)=0$, convexity gives $f_-\ge0$ on $[u_0,\infty)$, and concavity
then gives $f_-\ge0$ on $[0,u_0]$. Thus $\ln2\,\cH^2\le\ln2\,\Djs\le\cH^2$ cell by
cell. Disjoint supports attain the left constant; nearby distributions
approach the right one.
(ii) The records are affine in $\rho$, and the three divergences are jointly
convex; for $\cJ_\Sigma$ use convexity of mutual information in the channel.
Postprocessing is data processing.
(iii) If all records agree for every input, placing two measurements first
and exchanging them gives $\sqrt EF\sqrt E=\sqrt FE\sqrt F$ for every pair of effects of different measurements. This says that $\sqrt F\sqrt E$ is normal; its nonzero
spectrum equals that of $E^{1/4}\sqrt FE^{1/4}\succeq0$, so it is Hermitian
and the square roots commute~\cite{Gudder01}. The converse is immediate.
(iv) $\Tr(\sqrt{E_a}F_b\sqrt{E_a})=\Tr(E_aF_b)=\Tr(\sqrt{F_b}E_a\sqrt{F_b})$.
\emph{Relative entropy.} If $[P_a,Q_b]\ne0$, choose $b'\ne b$ and a unit vector
$\psi\in\ran Q_b$ with $Q_{b'}P_a\psi\ne0$. Then $p_{AB}(a,b')>0=p_{BA}(a,b')$.
\emph{Equation~\eqref{eq:Tspectral}.} $\TV(p,q)=\max_S\sum_{i\in S}(p_i-q_i)$;
exchange the maxima over $S$ and $\rho$ and use the variational
characterization of $\lambda_{\max}$.

\emph{Theorem~\ref{thm:binary}.} Expanding the projectors gives
$P_sQ_tP_s-Q_tP_sQ_t=\frac14(s\Lambda_1-t\Lambda_2)$ with $\Lambda_1=(A-BAB)/2=[A,B]B/2$ and
$\Lambda_2=(B-ABA)/2=[B,A]A/2$; the anticommutator terms cancel [this cancellation is the quantum question equality tested in
Ref.~\cite{Wang14}]. Summing absolute values over $s,t=\pm$ and using
$|u-v|+|u+v|=2\max(|u|,|v|)$ gives $\TV=\frac12\max(|\langle \Lambda_1\rangle|,|\langle
\Lambda_2\rangle|)$, and $\|\Lambda_1\|=\|\Lambda_2\|=\|[A,B]\|/2$ because $A$ and $B$ are unitary. On a
block with angle $\theta$, $\|[A,B]\|=2\sin\theta$. The record of any input is
the mixture, with the block populations as weights, of the block records
with common labels; off-diagonal blocks of $\rho$ do not contribute, and
one-dimensional blocks give equal records. Joint convexity and $D(p,p)=0$
show that the maximum is attained within one block. Replacing $\theta$ by
$\pi-\theta$ relabels the outcomes of $B$, so $\Psi_D$ depends only on
$\sin\theta$. The bounds on $\Psi$ follow from Eq.~\eqref{eq:JT} and
$1-h(\frac{1+u}2)\ge u^2/(2\ln2)$. For $\theta=\pi/2$, the records depend on
the Bloch components $(x,z)$ of the input in the plane of $A$ and $B$ and are
affine in them, so by joint convexity the maximum lies on the circle
$x^2+z^2=1$. The series $h(\frac{1+u}2)=1-\sum_{n\ge1}u^{2n}/[2n(2n-1)\ln2]$
gives $\Djs=\sum_n[\frac12(x^{2n}+z^{2n})-2^{-2n}\frac{(x+z)^{2n}+(x-z)^{2n}}2]
/[2n(2n-1)\ln2]$. On the circle $x^{2n}+z^{2n}\le1$ and
$\frac12[(x+z)^{2n}+(x-z)^{2n}]\ge(x^2+z^2)^n=1$ by convexity of
$t\mapsto t^n$, so each term is at most its value at $(1,0)$, and
$\Psi(1/2)=h(3/4)-\frac12$.

\emph{Theorem~\ref{thm:support}.} Put $\nu_{ab}=\langle P_a\psi,Q_b\psi\rangle$,
so that $\sum_{a,b}\nu_{ab}=1$ with nonzero terms only for occupied $a,b$.
Cauchy--Schwarz gives $|\nu_{ab}|\le[q_bp_{AB}(a,b)]^{1/2}$ and
$|\nu_{ab}|\le[p_ap_{BA}(a,b)]^{1/2}$. Taking the geometric mean of the two bounds with weights $z$ and $1-z$ gives
$|\nu_{ab}|\le[p_{AB}(a,b)^z\,p_{BA}(a,b)^{1-z}]^{1/2}\,[q_b^z\,p_a^{1-z}]^{1/2}$.
Summing over $a,b$ and applying the Cauchy--Schwarz inequality to the two
square roots gives
$1=|\sum_{ab}\nu_{ab}|\le[\chi_z(\sum_ap_a^{1-z})(\sum_bq_b^z)]^{1/2}$, and concavity of powers
gives the second inequality. With $z=1/2$, $1-\cH^2(A,B|\psi)=\chi_{1/2}$.
Since $r^{-z}s^{z-1}\ge1/\max(r,s)$, $\xi(\psi)\le\log_2\max(n,m)$. Equality $\chi_z=r_\psi^{-z}s_\psi^{z-1}$ forces uniform marginals, equal positive terms on
the occupied rectangle, a common phase of the $\nu_{ab}$, and equality in
both pointwise bounds, i.e., the two identities of the theorem. The normalized branch
vectors then form two orthonormal sets with cross Gram matrix of constant
entries $(r_\psi s_\psi)^{-1/2}$, whose only nonzero singular value is one; the
two spans meet exactly in $\mathbb C\psi$ and are otherwise orthogonal, so
$d\ge r_\psi+s_\psi-1$. The construction in the proof of Theorem~\ref{thm:separation}(b) realizes this
geometry for $r_\psi,s_\psi\ge2$ and $d=r_\psi+s_\psi-1$. Larger $d$ is
covered by assigning the extra orthogonal subspace to any projector of each
measurement, and $r_\psi=1$ by taking $P_1=\proj\psi$. For the exponent,
assume $2\le n\le m$, take an $m$-outcome Fourier PVM and a computational PVM grouped into $n$
sets, one of which is a singleton, and take a vector of the singleton as
input. The records are $\delta_{a1}/m$ and $k_a/m^2$, where $k_a$ is the size
of group $a$, so
$\chi_z=m^{z-1}\to1/m$ as $z\to0$.

\emph{Theorem~\ref{thm:separation}.} (a) The records are
$p_{AB}(a,b)=p_a/n$ and $p_{BA}(a,b)=q_b/n$, and $\|P_aQ_b\|^2=1/n$. Since
$|\nu_{ab}|\le\|P_aQ_b\|\,(p_aq_b)^{1/2}$, the identity $\sum\nu_{ab}=1$ gives
$(\sum_a\sqrt{p_a})(\sum_b\sqrt{q_b})\ge\sqrt n$, so the affinity
$(\sum_a\sqrt{p_a})(\sum_b\sqrt{q_b})/n$ is at least $n^{-1/2}$. The common mass, which equals $1-\TV$,
$\sum_{ab}\min(p_a,q_b)/n$ is at least $\sum_{ab}p_aq_b/n=1/n$. A unit vector in
$\ran P_a$ attains both bounds, the exponent $\log_2n$, and the
Maassen--Uffink bound $\Ustar\ge-\log_2c^2=\log_2n$. For MUBs, an eigenstate
gives $\Djs=\varphi(1/d)$. For the upper bound in Eq.~\eqref{eq:mubJ}, use
$\Djs\le\sum_i\frac{(x_i-y_i)^2}{2(x_i+y_i)}$, which follows from
$h(t)\ge4t(1-t)$, the collision bound $\sum_ap_a^2+\sum_bq_b^2\le1+1/d$ for
MUBs~\cite{WuYuMolmer09}. In detail, with $S_2=\sum_ap_a^2+\sum_bq_b^2$,
\[
 \Djs\le1-\frac2d\sum_{ab}\frac{p_aq_b}{p_a+q_b}
 \le1-\frac{2}{dS_2}\le\frac{d-1}{d+1},
\]
where zero cells are omitted and Cauchy--Schwarz gives
$\sum p_aq_b/(p_a+q_b)\ge(\sum p_aq_b)^2/\sum p_aq_b(p_a+q_b)=1/S_2$.
Equality would require $p_a=q_b$ on all occupied pairs and hence uniform
supports of size $2d/(d+1)\notin\mathbb Z$. The qubit equality is the case
$\theta=\pi/2$ of Theorem~\ref{thm:binary}.
(b) Let $F_t$ be the $t$-dimensional Fourier matrix, let $\tilde V_r$
consist of its last $r-1$ columns for $t=r$, and let $\tilde W_s$ consist of
the last $s-1$ rows of $F_s$. The matrix
\begin{equation}
 V=\begin{pmatrix}\mathbf 1_{r\times s}/\sqrt{rs}&\tilde V_r\\
 \tilde W_s&0\end{pmatrix}
\end{equation}
is unitary, because the nonconstant Fourier modes are orthogonal to the
constant vector. Take the computational basis as $A$, the columns of $V$ as
$B$, and $\psi=r^{-1/2}(1,\ldots,1,0,\ldots,0)$ with $r$ nonzero entries. On the
occupied rectangle the two records equal $1/(r^2s)$ and $1/(rs^2)$, and
elsewhere their supports are disjoint. Hence $\cH^2\ge1-(rs)^{-1/2}$, and for
$r=s$ the records coincide on a set of mass $1/r$ and are disjoint
elsewhere, so $\cJ\ge1-1/r=(n-1)/(n+1)$. The largest squared overlap is
$1/\min(r,s)$, attained in the off-diagonal blocks, and a basis vector of
such a block has a uniform distribution over $\min(r,s)$ outcomes of the other
basis, so it attains the Maassen--Uffink bound: $\Ustar=\log_2\min(r,s)$.
Tensoring all projectors with $\one_k$ leaves every record and every entropy
unchanged. Finally $rs-n=(r-1)(s-1)>0$.
(c) The off-diagonal blocks of $\rho$ do not affect the records or the
marginals. With block weights $w_j$ and normalized block states $\rho_j$, both
records of $\rho$ are the corresponding mixtures of block records. Joint
convexity bounds each divergence by the weighted average of the block
divergences, hence by their maximum, which is attained by an input in one
block. Concavity of the entropy gives $H(A)_\rho+H(B)_\rho\ge\sum_jw_j
[H(A)_{\rho_j}+H(B)_{\rho_j}]$, hence $\Ustar$ is at least the smallest block
value, which is again attained within one block.

\emph{Remark~\ref{rem:marginal}.} The marginal of the reversed record is the
disturbed marginal, so its distance to the undisturbed marginal is a
coarse-graining of $\TV(p_{AB},p_{BA})$. For a three-outcome measurement every
nontrivial event is a singleton or its complement, so the optimized marginal
disturbance is the largest $\|P_a-\sum_bQ_bP_aQ_b\|$ over $a$, together with the
same expression with $A$ and $B$ exchanged~\cite{Mandayam14}; $\cT$ is given by
Eq.~\eqref{eq:Tspectral}. Both are finite eigenvalue problems.

\emph{Equation~\eqref{eq:unsharp}.} Write
$a=\eta(1-\sqrt{1-\zeta^2})$ and
$b=\zeta(1-\sqrt{1-\eta^2})$.
Direct multiplication of the positive square roots of the two effects gives
\[
 X_{st}-Y_{st}=\tfrac14(sa\,\sigma_z-tb\,\sigma_x),\qquad s,t=\pm1.
\]
For an input with Bloch components $r_x,r_z$ this yields
$\TV=\frac12\max\{a|r_z|,b|r_x|\}$, using
$|u-v|+|u+v|=2\max(|u|,|v|)$. Maximizing over the Bloch ball proves
Eq.~\eqref{eq:unsharp}, including $\eta=0,1$ and $\zeta=0,1$.

\section{Proof of Theorem~\ref{thm:uncertainty}}\label{app:uncert}
(a) Choose $(a,b)$ with $\|P_aQ_b\|=c$ and a unit vector $\psi\in\ran P_a$ with
$\|Q_b\psi\|=c$. In the order $AB$ the outcome of $A$ is $a$ with certainty.
In the order $BA$ it is $a$ with probability
$s=\sum_{b'}\|P_aQ_{b'}\psi\|^2\le\sum_{b'}\|P_aQ_{b'}\|^2\|Q_{b'}\psi\|^2\le c^2$.
Coarse-graining both records to the event ``$A=a$'' gives $\TV\ge1-s\ge1-c^2$
and $\Djs\ge h(\frac{1+s}2)-\frac{h(s)}2=\varphi(s)\ge\varphi(c^2)$, since
$\varphi$ decreases on $[0,1]$. For MUBs $\cT=1-1/d$ by
Theorem~\ref{thm:separation}(a), and for qubit MUBs $\cJ=\varphi(1/2)$.
(b) The same $\psi$ has $H(A)=0$, and its $B$ distribution has largest weight
at least $c^2$. Fano's inequality~\cite{Cover06} gives
$H(B)\le h(1-q_{\max})+(1-q_{\max})\log_2(m-1)\le g_m(1-c^2)$, where
$c^2\ge\max_b\|Q_b\psi'\|^2\ge1/m$ for any unit $\psi'\in\ran P_a$ keeps the
argument in the range where $g_m$ increases. Exchanging the roles of $A$ and
$B$ gives $g_n$, and $g_k\le g_m,g_n$. Combine with (a). Finally
$\Ustar\le\log_2k$: an eigenvector of the measurement with more outcomes has
zero entropy for that measurement and at most $\log_2k$ for the other.
(c) $|\langle[A,B]\rangle|\le\|[A,B]\|=4\cT$, with equality on an extremal
eigenvector of $i[A,B]$. For the variance,
$\Var A+\Var B=2-\langle A\rangle^2-\langle B\rangle^2$. The pair of means of an
input is a convex combination of pairs of means of block states, so the
convex function $\langle A\rangle^2+\langle B\rangle^2$ attains its maximum
within one block, with value $2$ on a one-dimensional block and
$1+|\cos\theta_j|$ on a two-dimensional one. Thus
$U_{\rm var}=\min_j(1-|\cos\theta_j|)\le1-\sqrt{1-4\cT^2}$, with equality for a
single block.
(d) By the first inequality in Eq.~\eqref{eq:support} with $z=1/2$,
$H_{1/2}(A)+H_{1/2}(B)\ge-2\log_2\chi_{1/2}=-2\log_2(1-\cH^2(A,B|\rho))$ for pure
inputs; for mixed inputs apply this to a purification and to the PVMs
$P_a\otimes\one$ and $Q_b\otimes\one$. Then use $\cH^2\ge\ln2\,\cJ$
from Eq.~\eqref{eq:JH}.

\section{Proofs for Sec.~\ref{sec:sequential}}\label{app:sequential}
\emph{Lemma~\ref{lem:seq}.} (i) is trace preservation. (ii) For L\"uders instruments, commuting effects $E_1,E_2$ give
$\Tr[\sqrt{E_2}\sqrt{E_1}\rho\sqrt{E_1}\sqrt{E_2}]=\Tr(\rho E_1E_2)$, and
later measurements do not change the marginal. (iii) is data processing.

\emph{Theorem~\ref{thm:context}.} Since $p^\rho_\pi$ is a joint distribution,
the inequality gives $\sum_C\sum_x\alpha_C(x)\,p^\rho_\pi|_C(x)\le b$. For each
context, after shifting $\alpha_C$ by $(\max\alpha_C+\min\alpha_C)/2$, which
does not change $\sum_x\alpha_C(x)[p_C(x)-p^\rho_\pi|_C(x)]$,
\begin{align*}
 \textstyle\sum_x\alpha_C(x)[p^\rho_C(x)-p^\rho_\pi|_C(x)]
 &\le\textstyle\frac{\omega_C}2\sum_x|p^\rho_C(x)-p^\rho_\pi|_C(x)|\\
 &=\omega_C\TV(p^\rho_C,p^\rho_\pi|_C).
\end{align*}
Lemma~\ref{lem:seq}(ii) replaces $p^\rho_C$ by $p^\rho_{\pi_C}|_C$. The bound on
$\delta_{\rm NC}$ uses the joint distribution $p^\rho_\pi$, and the statement
on order information uses Eq.~\eqref{eq:JT}. The Peres--Mermin and KCBS minima were computed by depth-first enumeration of
schedules with directly evaluated sequential L\"uders records and pruning of
partial sums. The Peres--Mermin value was also checked over all $9!$
schedules using the Pauli commutation rule, and the KCBS value by direct
records for all $120$ schedules. These numerical evaluations use floating-point
arithmetic.

\emph{Exclusivity.} For a 0/1 assignment with support $S$, let $V_c$ be a
minimal vertex cover of the edges of $\mathcal G$ inside $S$. Minimality gives
every $v\in V_c$ an edge $e_v$ inside $S$ that is covered only by $v$, and
these edges are distinct. Since $c_{ij}\ge0$, dropping the other edges gives
$\sum_{i\in S}w_i-\sum_{ij\in E(\mathcal G[S])}c_{ij}\le
\sum_{i\in S\setminus V_c}w_i+\sum_{v\in V_c}(w_v-c_{e_v})\le
\alpha(\mathcal G,w)$, because $S\setminus V_c$ is independent and
$c_{e_v}\ge w_v$. On quantum behaviors the subtracted terms vanish, since
exclusive projectors are orthogonal.

\emph{Theorem~\ref{thm:spatial}.} We prove part (b) first, since its last
step gives the local-hidden-state model of part (a). (b) Local operations of
different parties commute. Moving setting $x$ to the front for party $k$
changes the joint record only through party $k$; conditioned on the outcomes
of the other parties, the change is a total-variation distance between two
local schedules on a conditional state of party $k$, which is at most $\cT_k(\pi_k,\pi_k^{x_k})$.
If all parties but one are scheduled, their records $\lambda$ and the
conditional states $\sigma_\lambda$ of the remaining party form a
local-hidden-state model, which is local for all its measurements. (a) For the assemblage,
$\sigma_{a|x}-\tau_{a|x}=\Tr_A[(\Delta_a\otimes\one)\rho]$ with
$\Delta_a=E_{a|x}-\mathcal E^*(E_{a|x})$, where $\mathcal E^*$ is the Heisenberg-picture map
of the settings that precede $x$ in $\pi$. Hence
$\|\sigma_{a|x}-\tau_{a|x}\|_1\le\|\Delta_a\|=\max_\sigma|\Tr\sigma\Delta_a|$, and $|\Tr\sigma\Delta_a|$ is at most the total-variation distance of the two
$x$-marginals on $\sigma$, hence at most $\cT_A(\pi,\pi^x)$. Summing over $a$ gives the factor
$n_x$. Finally, since $\sum_a(\sigma_{a|x}-\tau_{a|x})=0$,
$\sum_a\Tr[\Phi_{a|x}(\sigma_{a|x}-\tau_{a|x})]=\sum_a\Tr[(\Phi_{a|x}-\mu
\one)(\sigma_{a|x}-\tau_{a|x})]\le\max_a\|\Phi_{a|x}-\mu\one\|\,n_x\cT_A(\pi,\pi^x)$. For CHSH in part (b), scheduling Alice's settings as $(1,2)$ gives errors $0$ for $x=1$
and at most $\cT(A_1,A_2)$ for $x=2$. Each correlator changes by at most twice
the total-variation distance, so $\beta-2\le4\cT(A_1,A_2)$.

\section{Proofs for Sec.~\ref{sec:product}}\label{app:product}
\emph{Theorem~\ref{thm:product}.} (a) Write
$A_xA_{x'}=A_x\circ A_{x'}+iK^A_{xx'}$ and $B_yB_{y'}=B_y\circ B_{y'}+iK^B_{yy'}$
with the Hermitian operators $K=[\cdot,\cdot]/(2i)$. Then
$\mathfrak B^2=\sum M_{xy}M_{x'y'}(A_x\circ A_{x'}+iK^A_{xx'})\otimes(B_y\circ
B_{y'}+iK^B_{yy'})$. Under the relabeling $(x,y)\leftrightarrow(x',y')$ the
Jordan products are symmetric and the $K$'s antisymmetric, so the mixed terms
cancel, which gives Eq.~\eqref{eq:Bsquare}. Finally
$\|K^A_{xx'}\|=2\cT^A_{xx'}$ by Theorem~\ref{thm:binary}, and the terms with
$x=x'$ or $y=y'$ vanish.
(b) With the decomposition,
$\Gamma=\sum_{\bm s}\Omega_{\bm s}\otimes\sum_{x,x'}s_xs_{x'}(R_x\circ R_{x'})
=\sum_{\bm s}\Omega_{\bm s}\otimes(\sum_xs_xR_x)^2$. Each term is at most
$\|\sum_xs_xR_x\|^2\,\Omega_{\bm s}\otimes\one$, and
$\|\sum_xs_xR_x\|\le\sum_y|(M^{\mathsf T}\bm s)_y|\le\bC$. For two involutions
$\|A_1\circ A_2\|\le1$, so $\Omega_\pm\succeq0$. For commuting involutions $\Omega_{\bm s}$ is the sum of the joint spectral
projectors of $\bm s$ and $-\bm s$. For three involutions the
four sign patterns modulo a global sign and the four conditions
$\sum_{\bm s}\Omega_{\bm s}=\one$ and $\sum_{\bm s}s_xs_{x'}\Omega_{\bm s}=
A_x\circ A_{x'}$ determine $\Omega_{\bm s}=[(\sum_xs_xA_x)^2-\one]/8$ uniquely.
(c) Jordan's lemma decomposes the space of Alice into invariant blocks on
which $A_1\circ A_2=\cos\theta_j\one$ and $[A_1,A_2]/2=i\sin\theta_jY_j$, where
$Y_j$ is an involution on two-dimensional blocks and $\sin\theta_j=0$ on
one-dimensional ones; $\mathfrak B^2$ is block diagonal accordingly, and Bob's
operators act on his own factor. On a block, $\mathfrak B^2=R_1^2+R_2^2+\cos\theta_j\{R_1,R_2\}+\sin\theta_jY_j\otimes
i[R_1,R_2]$, which gives Eq.~\eqref{eq:twosetting}. Its Jordan part equals
$\frac{1+\cos\theta_j}2(R_1+R_2)^2+\frac{1-\cos\theta_j}2(R_1-R_2)^2\le
\max_\pm\|R_1\pm R_2\|^2\le\bC^2$. With $\sin\theta_j\le2\cT^A_{12}$ and
$\|[R_1,R_2]\|\le4\sum_{y\ne y'}|M_{1y}M_{2y'}|\cT^B_{yy'}$ the bound follows. For
CHSH, $R_1^2+R_2^2=4\one$, $\{R_1,R_2\}=0$, and $i[R_1,R_2]$ has symmetric
spectrum with norm $8\cT_B$, so equality holds.

\emph{Corollary~\ref{cor:DI}.} Rearrange part (c). For CHSH,
$\sum_{y\ne y'}|M_{1y}M_{2y'}|=2$. At $\beta=2\sqrt2$, $\cT_A\cT_B\ge1/4$ and
$\cT\le1/2$ force $\cT_A=\cT_B=1/2$; Theorem~\ref{thm:spatial}(b) gives
$\min\{\cT_A,\cT_B\}\ge(2\sqrt2-2)/4$.

\emph{Proposition~\ref{prop:HSG}.} (i) With $U$ and $V$ the matrices whose rows
are $\bm u_x$ and $\bm v_y$, $\|U^{\mathsf T}MV\|_F^2=\Tr(G_AMG_BM^{\mathsf T})$ with
the Gram matrices $G_A=UU^{\mathsf T}$ and $G_B=VV^{\mathsf T}$, and every
real Gram matrix of unit vectors is of this form. By duality, $\|U^{\mathsf T}MV\|_F=\max_{\|W\|_F\le1}\sum_{xy}M_{xy}\bm
u_x^{\mathsf T}W\bm v_y$. The set of local correlation matrices $\mathrm{conv}\{\bm s\bm
t^{\mathsf T}\}$ is the unit ball of the norm dual to $M\mapsto\bC(M)$, so by
the bipolar theorem $Z$ is local if and only if $\sum M_{xy}Z_{xy}\le\bC(M)$ for
all $M$. The product-input statement follows from the expression of
$\langle\Gamma\rangle$ in terms of $G_A$ and $G_B$ given before the
proposition. (ii) Write $\Tr(G_A\Xi)$ with $\Xi=MG_BM^{\mathsf T}\succeq0$.
Nesterov's theorem~\cite{Nesterov98}, a form of the little Grothendieck
inequality, bounds its maximum over such $G_A$ by $\frac\pi2\max_{\bm s}\bm s^{\mathsf T}\Xi\bm s$, and
$\bm s^{\mathsf T}\Xi\bm s=\|\sum_y(M^{\mathsf T}\bm s)_y\bm v_y\|^2\le\bC^2$.
For completeness, Gaussian hyperplane rounding of a Gram representation of
$G_A$ produces signs with
$\mathbb E[s_xs_{x'}]=(2/\pi)\arcsin((G_A)_{xx'})$.
The entrywise matrix $\arcsin[G_A]-G_A$ is positive semidefinite by the
nonnegative Taylor coefficients of $\arcsin t-t$ and the Schur product
theorem. Taking its trace against $\Xi\succeq0$ gives
$\max_{\bm s}\bm s^{\mathsf T}\Xi\bm s\ge(2/\pi)\Tr(G_A\Xi)$.
(iii) Let $n\le3$. The set $\mathrm{conv}\{\bm s\bm t^{\mathsf T}\}$ is the cut
polytope of the complete bipartite graph $K_{n,m}$ in $\pm1$ coordinates.
$K_{3,m}$ has no $K_5$ minor: two branch sets consisting of vertices on the
$m$-side only are not adjacent, so at least four of five branch sets would
need one of the three vertices on the $n$-side. By Barahona and
Mahjoub~\cite{Barahona86}, the cut polytope of a graph without $K_5$ minor is
described by the bounds $|Z_{xy}|\le1$ and the inequalities of chordless
cycles. The chordless cycles of $K_{n,m}$ have length four, and their
inequalities are the CHSH inequalities
$\pm Z_{xy}\pm Z_{xy'}\pm Z_{x'y}\pm Z_{x'y'}\le2$ with an odd number of minus
signs. For $Z=UWV^{\mathsf T}$ with $\|W\|_F\le1$,
$|Z_{xy}|\le\|W\|\le1$. Changing the signs of $\bm u_x$ or $\bm v_y$
reduces every CHSH inequality to
$\bm u_x^{\mathsf T}W(\bm v_y+\bm v_{y'})+\bm u_{x'}^{\mathsf T}W(\bm v_y-\bm v_{y'})
\le2$. The vectors $\bm v_y\pm\bm v_{y'}$ are orthogonal with norms
$2\cos\phi$ and $2\sin\phi$; with orthonormal $\bm e_1,\bm e_2$ along them the
left side is at most $2(\cos\phi\|W\bm e_1\|+\sin\phi\|W\bm e_2\|)\le
2(\|W\bm e_1\|^2+\|W\bm e_2\|^2)^{1/2}\le2\|W\|_F$.
(iv) For $r$-dimensional vectors,
$\sum M_{xy}\bm u_x\cdot\bm v_y=\Tr(U^{\mathsf T}MV)\le\sqrt r\,\|U^{\mathsf T}MV\|_F$.
(v) Planar measurements on such a state have correlations
$\bm a_x^{\mathsf T}\mathrm{diag}(c,s)\bm b_y$, and with vanishing local Bloch
vectors locality of the correlations is equivalent to locality of the full
behavior. They are local for $c^2+s^2\le1$ by Eq.~\eqref{eq:HSG}, and CHSH is
violated for suitable pairs of directions when $c^2+s^2>1$: the optimized
planar value is $2\sqrt{c^2+s^2}$~\cite{Horodecki95}. States with
$T=\mathrm{diag}(c,s,t)$ exist for every $c^2+s^2\le1$, e.g., Bell-diagonal
states with suitable $t$.

\emph{Theorem~\ref{thm:steer}.} With $\mathfrak S=\sum_xA_x\otimes B_x$ and
$B_{x'}B_x=-B_xB_{x'}$ for $x\ne x'$,
$\mathfrak S^2=n\one+\sum_{x<x'}[A_x,A_{x'}]\otimes B_xB_{x'}$. Each $B_xB_{x'}$ is
an anti-Hermitian unitary, so each term is Hermitian with norm
$\|[A_x,A_{x'}]\|=4\cT^A_{xx'}$. For a local-hidden-state model,
$\langle\mathfrak S\rangle=\sum_\lambda p_\lambda\sum_xa_x(\lambda)\langle B_x
\rangle_{\sigma_\lambda}$ with $|a_x(\lambda)|\le1$, and
$\sum_x\langle B_x\rangle^2_\sigma\le1$ because
$(\sum_xc_xB_x)^2=\|\bm c\|^2\one$; Cauchy--Schwarz gives $\sqrt n$. For $n=2$,
the Hermitian unitary $iB_1B_2$ anticommutes with $B_1$ and $B_2$, so
conjugation by $\one\otimes iB_1B_2$ maps $\mathfrak S$ to $-\mathfrak S$.
Conjugation by $A_1\otimes\one$ maps the commutator term to its negative, so its
spectrum is symmetric, $\lambda_{\max}(\mathfrak S^2)=2+4\cT^A_{12}$, and the
bound is attained. For Pauli
triples for both parties, $\mathfrak S$ is Bell diagonal with eigenvalues of
modulus $1$ and $3$, so $\|\mathfrak S\|=3=(3+4\cdot\frac32)^{1/2}$.

\emph{Corollary~\ref{cor:minimal}.} (i) is Theorem~\ref{thm:steer} with $n=2$.
(ii) For CHSH, Theorem~\ref{thm:product}(c) with Bob's commutator attaining
its norm on a product of extremal eigenvectors; with Bob optimal,
$\cT_B=1/2$. For MABK, put $\Theta_k=i[X_{k,1},X_{k,2}]/2$,
$S_n=\mathfrak M_n^2=(\mathfrak M'_n)^2$ and
$T_n=i[\mathfrak M_n,\mathfrak M'_n]/2$. Expanding the recursion proves
the equality of the two squares and gives
\begin{align}
S_n&=S_{n-1}\otimes\one+T_{n-1}\otimes\Theta_n,\nonumber\\
T_n&=T_{n-1}\otimes\one+S_{n-1}\otimes\Theta_n.\nonumber
\end{align}
Starting from $S_1=\one$ and $T_1=\Theta_1$, these are the even-
and odd-subset sums of the local $\Theta_k$, respectively. Thus
$\mathfrak M_N^2=\frac12[\bigotimes_k(\one+\Theta_k)+\bigotimes_k(\one-\Theta_k)]$.
For joint eigenvalues $t_k$ of the local $\Theta_k$, the corresponding
eigenvalue is $\sum_{S:\,|S|\ {\rm even}}\prod_{k\in S}t_k$.
Since $|t_k|\le2\cT_k$, it is at most the same sum with every
$t_k=2\cT_k$. Each $\Theta_k$ has symmetric spectrum, so these positive
extremal eigenvalues are simultaneously available on a product eigenvector.
This proves the formula; the two tensor products individually need not have
a common maximizing eigenvector. (iii) The bound
$\beta\le2\sqrt{1+C^2}$ holds for all two-qubit states~\cite{Verstraete02}, and
$\beta\le2\sqrt{1+4\cT_A\cT_B}$ by (ii). The first is attained for every $C$ by
a pure state with the Horodecki-optimal settings~\cite{Horodecki95}, the
second by a suitable maximally entangled state for every pair of local angles of nontrivial qubit observables.

\section{Proofs for Sec.~\ref{sec:three}}\label{app:three}
\emph{Theorem~\ref{thm:three}.} Let
$\mathcal L_Y(Z)=\frac12(Z+YZY)=(Y\circ Z)Y$ be the nonselective L\"uders map
of $Y$. It is unital and self-adjoint with respect to the trace inner product.
With $P_x=(\one+xX)/2$ one has $\sum_xxP_x\rho P_x=\frac12\{X,\rho\}$ and
$\sum_xP_x\rho P_x=\mathcal L_X(\rho)$. Applying these maps in the order of the
schedule $(X,Y,Z)$ and moving them to the observables gives, for the means in
the record,
\begin{equation}
 \begin{aligned}
 \mathbb E[xy]&=\langle X\circ Y\rangle,&
 \mathbb E[xz]&=\langle X\circ\mathcal L_Y(Z)\rangle,\\
 \mathbb E[yz]&=\langle\mathcal L_X(Y\circ Z)\rangle,&
 \mathbb E[xyz]&=\langle X\circ(Y\circ Z)\rangle .
 \end{aligned}
 \label{eq:moments}
\end{equation}
(a) For $\rho=\one/d$ the maps $\mathcal L$ leave the input invariant. The
means of single outcomes and of adjacent pairs are therefore $\Tr(X)/d$,
$\Tr(XY)/d$, and $\Tr(YZ)/d$, and $\mathbb E[xyz]=\Tr(XYZ+XZY)/(2d)$ is
invariant under all permutations. For the outer pair,
$d\,\mathbb E[xz]=\Tr[X\mathcal L_Y(Z)]=\Tr[\mathcal L_Y(X)\mathcal L_Y(Z)]
=\Tr[(X\circ Y)(Y\circ Z)]=\Tr(XZ)+d\,\kappa_Y$. The steps use that $\mathcal L_Y$ is a self-adjoint projection, that
$Y\circ Z$ commutes with $Y$, and Eq.~\eqref{eq:compose}. Hence the record
depends only on $Y$, and $8(p_{(B)}-p_{(C)})=ac\,\kappa_B-ab\,\kappa_C$, so that
$\TV(p_{(B)},p_{(C)})=\frac1{16}\sum_{a,b,c}|c\,\kappa_B-b\,\kappa_C|
=\frac14(|\kappa_B-\kappa_C|+|\kappa_B+\kappa_C|)$; the other pairs follow by
relabeling. The bound on $\kappa_B$ is the Cauchy--Schwarz inequality together
with $\|K\|_F\le\sqrt d\,\|K\|$.
(b) For signs, $(1-ab)(1-bc)=1-ab-bc+ac$ equals $4$ on the two alternating
strings and $0$ otherwise. Since
$\langle\Phi|X\otimes Y^{\mathsf T}|\Phi\rangle=\Tr(XY)/d$, the mean of this
function in $p_{(B)}$ is
$1-E_{AB}-E_{BC}+E_{AC}+\kappa_B$ by (a). For qubits,
$X\circ Y=(\bm x\cdot\bm y)\one$, $K_{XY}=(\bm x\times\bm y)\cdot\bm\sigma$, and
$\beta_3=(3-|\bm a-\bm b+\bm c|^2)/2$.

\emph{General inputs.} The Hermitian part of $AC=(AB)(BC)$ gives
\begin{multline}
 A\circ B+B\circ C-A\circ C-\one=K_{AB}\circ K_{BC}\\
 -(\one-A\circ B)\circ(\one-B\circ C)
 -\tfrac i2\big([A\circ B,K_{BC}]+[K_{AB},B\circ C]\big).
\end{multline}
The last term is odd under conjugation by $B$ and does not vanish for
general inputs. On $\mathbb C^3$ let
$A=\mathrm{diag}(1,-1,1)$, $B=\mathrm{diag}(-1,1,1)$,
$C=\one-2\bm w\bm w^{\mathsf T}$ with $\bm w=(\sqrt3,0,1)/2$, and
$\psi=(1,0,\sqrt3)/2$. Then $[A,B]=0$ and
$\langle A\circ B\rangle=\langle B\circ C\rangle=-\langle A\circ C\rangle=1/2$ on
$\psi$.

\emph{Theorem~\ref{thm:kcbs}.} (a) The vectors $v_i$ and $v_{i+2}$ lie in the
plane orthogonal to $v_{i+1}$. Let $u$ be the unit vector of this plane
orthogonal to $v_{i+2}$, so that $|\langle u,v_i\rangle|^2=1-z_i$. Since
$v_{i+3}\perp v_{i+2}$, $v_{i+3}=\alpha u+\beta v_{i+1}$ with
$|\beta|^2=z_{i+1}$, and
$z_{i+3}=|\langle v_{i+3},v_i\rangle|^2=|\alpha|^2(1-z_i)=(1-z_{i+1})(1-z_i)$.

(b) Let $\delta=2-\sum_iz_i$. We first show that
\begin{equation}
 \delta\,(1-z_{i+3})=z_iz_{i+1}z_{i+3}=(\cT_i\cT_{i+1})^2
 \quad\text{for all }i.
 \label{eq:deltachord}
\end{equation}
The second equality is (a). If some $z_j=1$, then $v_{j+2}$ is parallel to
$v_j$, the other three vectors lie in the plane orthogonal to $v_j$, and one
checks directly that $\delta=0$ and that both sides vanish. Otherwise (a),
applied at $i+2$ and $i+3$, gives $1-z_{i+2}=z_i/(1-z_{i+3})$ and
$1-z_{i+4}=z_{i+1}/(1-z_{i+3})$, hence
$\delta=(z_i+z_{i+1})z_{i+3}/(1-z_{i+3})-z_{i+3}=z_iz_{i+1}z_{i+3}/(1-z_{i+3})$,
where (a) was used once more. In particular $\delta\ge0$. For
$\Pi=\sum_iP_i$ one finds $\Tr\Pi=5$, $\Tr\Pi^2=5+2\sum_iz_i$, and
$\Tr\Pi^3=5+6\sum_iz_i$, because three vertices never form a triangle of
chords. Hence
$\det(\lambda-\Pi)=\lambda^3-5\lambda^2+(8+\delta)\lambda-4-3\delta$~\cite{Badziag11},
and $\lambda=2+V'$ gives the cubic $V'^3+V'^2+\delta V'-\delta$, whose roots
are real and have sum $-1$ and product $\delta$. For $\delta>0$ exactly one
root is positive; it lies in $(0,1)$ because the cubic is $-\delta$ at $0$ and
$2$ at $1$. For $\delta=0$ the largest root is $0$. The largest root is $V$,
and it satisfies Eq.~\eqref{eq:kcbsvalue}.

(c) The formula for $\cT_i\cT_{i+1}$ combines Eqs.~\eqref{eq:deltachord}
and~\eqref{eq:kcbsvalue}; it shows that $V\le\cT_i\cT_{i+1}$ if and only if
$z_{i+3}\le2V/(1+V)$. Let $V>0$ and let $\mu=z_k$ be the smallest overlap;
$0<\mu<1$ by Eq.~\eqref{eq:deltachord}. By (a), $\mu=(1-y)(1-y')$ with
$y=z_{k+2}$ and $y'=z_{k+3}$, and $y,y'\ge\mu$ gives $\mu\le(1-\mu)^2$, i.e.,
$\mu\le(3-\sqrt5)/2$. The sum of two numbers with product $\mu$ that do not
exceed $1-\mu$ is at most $1-\mu+\mu/(1-\mu)$; applied to $1-y$ and $1-y'$
this gives $yy'\ge\mu(1-2\mu)/(1-\mu)$, and Eq.~\eqref{eq:deltachord} with
$i=k+2$ gives $\delta=yy'\mu/(1-\mu)\ge\mu^2(1-2\mu)/(1-\mu)^2$. The right-hand
side of Eq.~\eqref{eq:kcbsvalue} increases with $V$ and equals
$\mu^2/[(2-\mu)^2(1-\mu)]$ at $V=\mu/(2-\mu)$. The inequality
$\mu^2(1-2\mu)/(1-\mu)^2\ge\mu^2/[(2-\mu)^2(1-\mu)]$ is equivalent to
$(\mu^2-3\mu+1)(3-2\mu)\ge0$, which holds. Hence $V\ge\mu/(2-\mu)$, i.e.,
$z_k\le2V/(1+V)$, and $V\le\cT_{k+2}\cT_{k+3}$. Equality requires
$\mu=(3-\sqrt5)/2$, hence $V=\sqrt5-2$, the largest possible
value~\cite{KCBS08}; then $\sum_iz_i=5\mu$ by Eq.~\eqref{eq:kcbsvalue}, so all
$z_i$ equal $\mu$.

\emph{Proposition~\ref{prop:cross}.} Chords $i$ and $j$ cross if and only if
$j-i\equiv\pm1$ modulo $5$, so any three chords contain a crossing pair.
(a) Let $j=5$ and $Q_i=\one-P_i$. Then
$-S-3\,\one=4(\sum_iP_i-\sum_iP_iP_{i+1}-2\,\one)$, and the eigenspaces of
$P_5$ are invariant. Since neighbors commute, the bracket equals
$P_1Q_2+P_4Q_3-Q_2Q_3-\one\preceq Q_2+Q_3-Q_2Q_3-\one=-P_2P_3\preceq0$ on the
eigenspace $P_5=0$ and
$P_2Q_1+P_3Q_2-P_3P_4-\one\preceq P_2+P_3Q_2-\one=-Q_2Q_3\preceq0$ on the
eigenspace $P_5=\one$. An involution $A_j$ commutes with all $A_i$ exactly
when both chords at the vertex $j$ commute, and two chords cannot meet all
five vertices; hence a violation requires three noncommuting chords.
(b) Let $[P_3,P_5]=0$, $E=P_3P_5$, and $\Pi=\sum_iP_i$. Exclusivity gives
$\Pi E=2E$, so a unit eigenvector $\psi$ of $\Pi$ with eigenvalue
$\lambda>2$ has $E\psi=0$. The vectors $P_i\psi$ are then orthogonal for
neighbors and for the chord $(3,5)$, so the Gram matrix of the normalized
nonzero ones is $\one+N$ with $N$ supported on a path. The spectrum of $N$ is
symmetric and $\one+N\succeq0$, hence $\one+N\preceq2\,\one$, and
$\lambda^2=\|\sum_iP_i\psi\|^2\le2\sum_i\|P_i\psi\|^2=2\lambda$, a
contradiction.

\section{Numerical methods}\label{app:numerics}
Maxima over inputs were computed by multistart local optimization (L-BFGS
followed by Nelder--Mead) over pure states; reported values are therefore
lower bounds on the maxima, and minimal entropy sums are upper bounds.
Exact identities and closed-form values were checked against direct matrix
computations for random involutions: Theorem~\ref{thm:binary} for
$d\le4$; the identity~\eqref{eq:Bsquare} and the two-setting bound for
$d_A,d_B\in\{2,3\}$ with up to four settings; the cut decomposition of three
involutions and Theorem~\ref{thm:steer} for $d_A\le4$;
Corollary~\ref{cor:minimal}(i),(ii) for $d_A\le5$ and up to four parties;
Eq.~\eqref{eq:unsharp} for five pairs $(\eta,\zeta)$;
Theorem~\ref{thm:three} and Appendix~\ref{app:three} for $500$ triples with $d\le6$;
Theorem~\ref{thm:kcbs} for approximately $4\times10^5$ random and $31$ degenerate
pentagons, with Eq.~\eqref{eq:kcbslaw} tested for $V>10^{-4}$;
Proposition~\ref{prop:cross} for $900$ instances. Locality of correlation
matrices was tested with finite-strategy linear programs solved by SciPy/HiGHS
in floating-point arithmetic, with success and residual checks. The word
``exact'' for Eq.~\eqref{eq:Tspectral} refers to exhaustive event enumeration,
not exact-arithmetic eigensolves. The Supplemental Material~\cite{SM}
provides reproducible checks of the numerical statements, the searches quoted
for Conjectures~\ref{conj:product} and~\ref{conj:kcbs}, and
Fig.~\ref{fig:minimal}. 

\bibliography{references}

@article{Heinosaari16,
 author={Heinosaari, Teiko and Miyadera, Takayuki and Ziman, M{\'a}rio},
 title={An invitation to quantum incompatibility},
 journal={J. Phys. A}, volume={49}, pages={123001}, year={2016},
 doi={10.1088/1751-8113/49/12/123001}}

@article{Guhne23RMP,
 author={G{\"u}hne, Otfried and Haapasalo, Erkka and Kraft, Tristan and Pellonp{\"a}{\"a}, Juha-Pekka and Uola, Roope},
 title={Colloquium: Incompatible measurements in quantum information science},
 journal={Rev. Mod. Phys.}, volume={95}, pages={011003}, year={2023},
 doi={10.1103/RevModPhys.95.011003}}

@article{Mandayam14,
 author={Mandayam, Prabha and Srinivas, M. D.},
 title={Measures of disturbance and incompatibility for quantum measurements},
 journal={Phys. Rev. A}, volume={89}, pages={062112}, year={2014},
 doi={10.1103/PhysRevA.89.062112}}

@article{Gao23,
 author={Gao, Ning and Li, Dantong and Mishra, Anchit and Yan, Junchen and Simonov, Kyrylo and Chiribella, Giulio},
 title={Measuring incompatibility and clustering quantum observables with a quantum switch},
 journal={Phys. Rev. Lett.}, volume={130}, pages={170201}, year={2023},
 doi={10.1103/PhysRevLett.130.170201}}

@article{Storrer26,
 author={Storrer, M. and Lima, P. and Costa, A. C. S. and P{\'a}dua, S. and Angelo, R. M.},
 title={Theory-independent context incompatibility: Quantification and experimental demonstration},
 journal={Phys. Rev. A}, volume={113}, pages={L010202}, year={2026},
 doi={10.1103/r2lr-72l7}}

@article{Atmanspacher12,
 author={Atmanspacher, Harald and R{\"o}mer, Hartmann},
 title={Order effects in sequential measurements of non-commuting psychological observables},
 journal={J. Math. Psychol.}, volume={56}, pages={274--280}, year={2012},
 doi={10.1016/j.jmp.2012.06.003}}

@article{Wang14,
 author={Wang, Zheng and Solloway, Tyler and Shiffrin, Richard M. and Busemeyer, Jerome R.},
 title={Context effects produced by question orders reveal quantum nature of human judgments},
 journal={Proc. Natl. Acad. Sci. USA}, volume={111}, pages={9431--9436}, year={2014},
 doi={10.1073/pnas.1407756111}}

@article{Gudder01,
 author={Gudder, Stanley and Nagy, Gabriel},
 title={Sequential quantum measurements},
 journal={J. Math. Phys.}, volume={42}, pages={5212--5222}, year={2001},
 doi={10.1063/1.1407837}}

@article{Bandyopadhyay13,
 author={Bandyopadhyay, Somshubhro and Mandayam, Prabha},
 title={Operational measure of incompatibility of noncommuting observables},
 journal={Phys. Rev. A}, volume={87}, pages={042120}, year={2013},
 doi={10.1103/PhysRevA.87.042120}}

@article{Landau87,
 author={Landau, L. J.},
 title={On the violation of {Bell's} inequality in quantum theory},
 journal={Phys. Lett. A}, volume={120}, pages={54--56}, year={1987},
 doi={10.1016/0375-9601(87)90075-2}}

@article{Kiukas10,
 author={Kiukas, Jukka and Werner, Reinhard F.},
 title={Maximal violation of {Bell} inequalities by position measurements},
 journal={J. Math. Phys.}, volume={51}, pages={072105}, year={2010},
 doi={10.1063/1.3447736}}

@article{Cereceda01,
 author={Cereceda, Jos{\'e} L.},
 title={Mermin's $n$-particle {Bell} inequality and operators' noncommutativity},
 journal={Phys. Lett. A}, volume={286}, pages={376--382}, year={2001},
 doi={10.1016/S0375-9601(01)00454-6}}

@article{Wolf09,
 author={Wolf, Michael M. and P{\'e}rez-Garc{\'i}a, David and Fern{\'a}ndez, Carlos},
 title={Measurements incompatible in quantum theory cannot be measured jointly in any other no-signaling theory},
 journal={Phys. Rev. Lett.}, volume={103}, pages={230402}, year={2009},
 doi={10.1103/PhysRevLett.103.230402}}

@article{Quintino14,
 author={Quintino, Marco T{\'u}lio and V{\'e}rtesi, Tam{\'a}s and Brunner, Nicolas},
 title={Joint measurability, {Einstein-Podolsky-Rosen} steering, and {Bell} nonlocality},
 journal={Phys. Rev. Lett.}, volume={113}, pages={160402}, year={2014},
 doi={10.1103/PhysRevLett.113.160402}}

@article{Uola14,
 author={Uola, Roope and Moroder, Tobias and G{\"u}hne, Otfried},
 title={Joint measurability of generalized measurements implies classicality},
 journal={Phys. Rev. Lett.}, volume={113}, pages={160403}, year={2014},
 doi={10.1103/PhysRevLett.113.160403}}

@article{Cavalcanti16,
 author={Cavalcanti, Daniel and Skrzypczyk, Paul},
 title={Quantitative relations between measurement incompatibility, quantum steering, and nonlocality},
 journal={Phys. Rev. A}, volume={93}, pages={052112}, year={2016},
 doi={10.1103/PhysRevA.93.052112}}

@article{Chen16,
 author={Chen, Shin-Liang and Budroni, Costantino and Liang, Yeong-Cherng and Chen, Yueh-Nan},
 title={Natural framework for device-independent quantification of quantum steerability, measurement incompatibility, and self-testing},
 journal={Phys. Rev. Lett.}, volume={116}, pages={240401}, year={2016},
 doi={10.1103/PhysRevLett.116.240401}}

@article{Guhne10,
 author={G{\"u}hne, Otfried and Kleinmann, Matthias and Cabello, Ad{\'a}n and Larsson, Jan-{\AA}ke and Kirchmair, Gerhard and Z{\"a}hringer, Florian and Gerritsma, Rene and Roos, Christian F.},
 title={Compatibility and noncontextuality for sequential measurements},
 journal={Phys. Rev. A}, volume={81}, pages={022121}, year={2010},
 doi={10.1103/PhysRevA.81.022121}}

@article{Szangolies13,
 author={Szangolies, Jochen and Kleinmann, Matthias and G{\"u}hne, Otfried},
 title={Tests against noncontextual models with measurement disturbances},
 journal={Phys. Rev. A}, volume={87}, pages={050101}, year={2013},
 doi={10.1103/PhysRevA.87.050101}}

@article{Kujala15,
 author={Kujala, Janne V. and Dzhafarov, Ehtibar N. and Larsson, Jan-{\AA}ke},
 title={Necessary and sufficient conditions for an extended noncontextuality in a broad class of quantum mechanical systems},
 journal={Phys. Rev. Lett.}, volume={115}, pages={150401}, year={2015},
 doi={10.1103/PhysRevLett.115.150401}}

@article{Budroni22,
 author={Budroni, Costantino and Cabello, Ad{\'a}n and G{\"u}hne, Otfried and Kleinmann, Matthias and Larsson, Jan-{\AA}ke},
 title={Kochen-{Specker} contextuality},
 journal={Rev. Mod. Phys.}, volume={94}, pages={045007}, year={2022},
 doi={10.1103/RevModPhys.94.045007}}

@article{Abramsky17,
 author={Abramsky, Samson and Barbosa, Rui Soares and Mansfield, Shane},
 title={Contextual fraction as a measure of contextuality},
 journal={Phys. Rev. Lett.}, volume={119}, pages={050504}, year={2017},
 doi={10.1103/PhysRevLett.119.050504}}

@article{Fine82,
 author={Fine, Arthur},
 title={Hidden variables, joint probability, and the {Bell} inequalities},
 journal={Phys. Rev. Lett.}, volume={48}, pages={291--295}, year={1982},
 doi={10.1103/PhysRevLett.48.291}}

@article{Peres90,
 author={Peres, Asher},
 title={Incompatible results of quantum measurements},
 journal={Phys. Lett. A}, volume={151}, pages={107--108}, year={1990},
 doi={10.1016/0375-9601(90)90172-K}}

@article{Mermin90PM,
 author={Mermin, N. David},
 title={Simple unified form for the major no-hidden-variables theorems},
 journal={Phys. Rev. Lett.}, volume={65}, pages={3373--3376}, year={1990},
 doi={10.1103/PhysRevLett.65.3373}}

@article{KCBS08,
 author={Klyachko, Alexander A. and Can, M. Ali and Binicio{\u{g}}lu, Sinem and Shumovsky, Alexander S.},
 title={Simple test for hidden variables in spin-1 systems},
 journal={Phys. Rev. Lett.}, volume={101}, pages={020403}, year={2008},
 doi={10.1103/PhysRevLett.101.020403}}

@article{WernerWolf01,
 author={Werner, Reinhard F. and Wolf, Michael M.},
 title={All-multipartite {Bell}-correlation inequalities for two dichotomic observables per site},
 journal={Phys. Rev. A}, volume={64}, pages={032112}, year={2001},
 doi={10.1103/PhysRevA.64.032112}}

@article{Cavalcanti09,
 author={Cavalcanti, E. G. and Jones, S. J. and Wiseman, H. M. and Reid, M. D.},
 title={Experimental criteria for steering and the {Einstein-Podolsky-Rosen} paradox},
 journal={Phys. Rev. A}, volume={80}, pages={032112}, year={2009},
 doi={10.1103/PhysRevA.80.032112}}

@article{Verstraete02,
 author={Verstraete, Frank and Wolf, Michael M.},
 title={Entanglement versus {Bell} violations and their behavior under local filtering operations},
 journal={Phys. Rev. Lett.}, volume={89}, pages={170401}, year={2002},
 doi={10.1103/PhysRevLett.89.170401}}

@article{Horodecki95,
 author={Horodecki, R. and Horodecki, P. and Horodecki, M.},
 title={Violating {Bell} inequality by mixed spin-1/2 states: necessary and sufficient condition},
 journal={Phys. Lett. A}, volume={200}, pages={340--344}, year={1995},
 doi={10.1016/0375-9601(95)00214-N}}

@article{Wootters98,
 author={Wootters, William K.},
 title={Entanglement of formation of an arbitrary state of two qubits},
 journal={Phys. Rev. Lett.}, volume={80}, pages={2245--2248}, year={1998},
 doi={10.1103/PhysRevLett.80.2245}}

@article{Robertson29,
 author={Robertson, H. P.},
 title={The uncertainty principle},
 journal={Phys. Rev.}, volume={34}, pages={163--164}, year={1929},
 doi={10.1103/PhysRev.34.163}}

@article{MU88,
 author={Maassen, Hans and Uffink, J. B. M.},
 title={Generalized entropic uncertainty relations},
 journal={Phys. Rev. Lett.}, volume={60}, pages={1103--1106}, year={1988},
 doi={10.1103/PhysRevLett.60.1103}}

@article{Krishna02,
 author={Krishna, M. and Parthasarathy, K. R.},
 title={An entropic uncertainty principle for quantum measurements},
 journal={Sankhy{\=a} A}, volume={64}, pages={842--851}, year={2002},
 note={arXiv:quant-ph/0110025}}

@article{Coles17,
 author={Coles, Patrick J. and Berta, Mario and Tomamichel, Marco and Wehner, Stephanie},
 title={Entropic uncertainty relations and their applications},
 journal={Rev. Mod. Phys.}, volume={89}, pages={015002}, year={2017},
 doi={10.1103/RevModPhys.89.015002}}

@article{Buscemi14,
 author={Buscemi, Francesco and Hall, Michael J. W. and Ozawa, Masanao and Wilde, Mark M.},
 title={Noise and disturbance in quantum measurements: An information-theoretic approach},
 journal={Phys. Rev. Lett.}, volume={112}, pages={050401}, year={2014},
 doi={10.1103/PhysRevLett.112.050401}}

@article{Abbott16,
 author={Abbott, Alastair A. and Alzieu, Pierre-Louis and Hall, Michael J. W. and Branciard, Cyril},
 title={Tight state-independent uncertainty relations for qubits},
 journal={Mathematics}, volume={4}, pages={8}, year={2016},
 doi={10.3390/math4010008}}

@article{Lin91,
 author={Lin, Jianhua},
 title={Divergence measures based on the {Shannon} entropy},
 journal={IEEE Trans. Inf. Theory}, volume={37}, pages={145--151}, year={1991},
 doi={10.1109/18.61115}}

@article{Chernoff52,
 author={Chernoff, Herman},
 title={A measure of asymptotic efficiency for tests of a hypothesis based on the sum of observations},
 journal={Ann. Math. Stat.}, volume={23}, pages={493--507}, year={1952},
 doi={10.1214/aoms/1177729330}}

@article{Tavakoli21,
 author={Tavakoli, Armin and Farkas, M{\'a}t{\'e} and Rosset, Denis and Bancal, Jean-Daniel and Kaniewski, J{\k{e}}drzej},
 title={Mutually unbiased bases and symmetric informationally complete measurements in {Bell} experiments},
 journal={Sci. Adv.}, volume={7}, pages={eabc3847}, year={2021},
 doi={10.1126/sciadv.abc3847}}

@article{Farkas23,
 author={Farkas, M{\'a}t{\'e} and Kaniewski, J{\k{e}}drzej and Nayak, Ashwin},
 title={Mutually unbiased measurements, {Hadamard} matrices, and superdense coding},
 journal={IEEE Trans. Inf. Theory}, volume={69}, pages={3814--3824}, year={2023},
 doi={10.1109/TIT.2023.3269524}}

@article{Halmos69,
 author={Halmos, P. R.},
 title={Two subspaces},
 journal={Trans. Amer. Math. Soc.}, volume={144}, pages={381--389}, year={1969},
 doi={10.1090/S0002-9947-1969-0251519-5}}

@article{Yu10,
 author={Yu, Sixia and Liu, Nai-le and Li, Li and Oh, C. H.},
 title={Joint measurement of two unsharp observables of a qubit},
 journal={Phys. Rev. A}, volume={81}, pages={062116}, year={2010},
 doi={10.1103/PhysRevA.81.062116}}

@book{Cover06,
 author={Cover, Thomas M. and Thomas, Joy A.},
 title={Elements of Information Theory}, edition={2nd},
 publisher={Wiley}, address={Hoboken, NJ}, year={2006}}

@article{Uola15,
 author={Uola, Roope and Budroni, Costantino and G{\"u}hne, Otfried and Pellonp{\"a}{\"a}, Juha-Pekka},
 title={One-to-one mapping between steering and joint measurability problems},
 journal={Phys. Rev. Lett.}, volume={115}, pages={230402}, year={2015},
 doi={10.1103/PhysRevLett.115.230402}}

@article{Krivine79,
 author={Krivine, J. L.},
 title={Constantes de {Grothendieck} et fonctions de type positif sur les sph{\`e}res},
 journal={Adv. Math.}, volume={31}, pages={16--30}, year={1979},
 doi={10.1016/0001-8708(79)90017-3}}

@article{AcinGisinToner06,
 author={Ac{\'i}n, Antonio and Gisin, Nicolas and Toner, Benjamin},
 title={{Grothendieck's} constant and local models for noisy entangled quantum states},
 journal={Phys. Rev. A}, volume={73}, pages={062105}, year={2006},
 doi={10.1103/PhysRevA.73.062105}}

@article{Nesterov98,
 author={Nesterov, Yu.},
 title={Semidefinite relaxation and nonconvex quadratic optimization},
 journal={Optim. Methods Softw.}, volume={9}, pages={141--160}, year={1998},
 doi={10.1080/10556789808805690}}

@article{Tsirelson80,
 author={Cirel'son, B. S.},
 title={Quantum generalizations of {Bell's} inequality},
 journal={Lett. Math. Phys.}, volume={4}, pages={93--100}, year={1980},
 doi={10.1007/BF00417500}}

@article{Tsirelson87,
 author={Tsirel'son, B. S.},
 title={Quantum analogues of the {Bell} inequalities. {The} case of two spatially separated domains},
 journal={J. Sov. Math.}, volume={36}, pages={557--570}, year={1987},
 doi={10.1007/BF01663472}}

@article{OppenheimWehner10,
 author={Oppenheim, Jonathan and Wehner, Stephanie},
 title={The uncertainty principle determines the nonlocality of quantum mechanics},
 journal={Science}, volume={330}, pages={1072--1074}, year={2010},
 doi={10.1126/science.1192065}}

@article{HeinosaariWolf10,
 author={Heinosaari, Teiko and Wolf, Michael M.},
 title={Nondisturbing quantum measurements},
 journal={J. Math. Phys.}, volume={51}, pages={092201}, year={2010},
 doi={10.1063/1.3480658}}

@article{Andersson05,
 author={Andersson, Erika and Barnett, Stephen M. and Aspect, Alain},
 title={Joint measurements of spin, operational locality, and uncertainty},
 journal={Phys. Rev. A}, volume={72}, pages={042104}, year={2005},
 doi={10.1103/PhysRevA.72.042104}}

@article{Busch86,
 author={Busch, Paul},
 title={Unsharp reality and joint measurements for spin observables},
 journal={Phys. Rev. D}, volume={33}, pages={2253--2261}, year={1986},
 doi={10.1103/PhysRevD.33.2253}}

@article{Briet14,
 author={Bri{\"e}t, Jop and de Oliveira Filho, Fernando M{\'a}rio and Vallentin, Frank},
 title={{Grothendieck} inequalities for semidefinite programs with rank constraint},
 journal={Theory Comput.}, volume={10}, pages={77--105}, year={2014},
 doi={10.4086/toc.2014.v010a004}}

@article{Hirsch17,
 author={Hirsch, Flavien and Quintino, Marco T{\'u}lio and V{\'e}rtesi, Tam{\'a}s and Navascu{\'e}s, Miguel and Brunner, Nicolas},
 title={Better local hidden variable models for two-qubit {Werner} states and an upper bound on the {Grothendieck} constant {$K_G(3)$}},
 journal={Quantum}, volume={1}, pages={3}, year={2017},
 doi={10.22331/q-2017-04-25-3}}

@article{Designolle19,
 author={Designolle, S{\'e}bastien and Farkas, M{\'a}t{\'e} and Kaniewski, J{\k{e}}drzej},
 title={Incompatibility robustness of quantum measurements: a unified framework},
 journal={New J. Phys.}, volume={21}, pages={113053}, year={2019},
 doi={10.1088/1367-2630/ab5020}}

@article{Designolle23,
 author={Designolle, S{\'e}bastien and Iommazzo, Gabriele and Besan{\c{c}}on, Mathieu and Knebel, Sebastian and Gel{\ss}, Patrick and Pokutta, Sebastian},
 title={Improved local models and new {Bell} inequalities via {Frank-Wolfe} algorithms},
 journal={Phys. Rev. Research}, volume={5}, pages={043059}, year={2023},
 doi={10.1103/PhysRevResearch.5.043059}}

@article{Kunjwal14,
 author={Kunjwal, Ravi and Heunen, Chris and Fritz, Tobias},
 title={Quantum realization of arbitrary joint measurability structures},
 journal={Phys. Rev. A}, volume={89}, pages={052126}, year={2014},
 doi={10.1103/PhysRevA.89.052126}}

@article{Fritz10,
 author={Fritz, Tobias},
 title={Quantum correlations in the temporal {Clauser-Horne-Shimony-Holt} ({CHSH}) scenario},
 journal={New J. Phys.}, volume={12}, pages={083055}, year={2010},
 doi={10.1088/1367-2630/12/8/083055}}

@article{Budroni13,
 author={Budroni, Costantino and Moroder, Tobias and Kleinmann, Matthias and G{\"u}hne, Otfried},
 title={Bounding temporal quantum correlations},
 journal={Phys. Rev. Lett.}, volume={111}, pages={020403}, year={2013},
 doi={10.1103/PhysRevLett.111.020403}}

@article{Heinosaari15,
 author={Heinosaari, Teiko and Kiukas, Jukka and Reitzner, Daniel},
 title={Noise robustness of the incompatibility of quantum measurements},
 journal={Phys. Rev. A}, volume={92}, pages={022115}, year={2015},
 doi={10.1103/PhysRevA.92.022115}}

@article{Barahona86,
 author={Barahona, Francisco and Mahjoub, Ali Ridha},
 title={On the cut polytope},
 journal={Math. Program.}, volume={36}, pages={157--173}, year={1986},
 doi={10.1007/BF02592023}}

@article{Schneeloch13,
 author={Schneeloch, James and Broadbent, Curtis J. and Walborn, Stephen P. and Cavalcanti, Eric G. and Howell, John C.},
 title={{Einstein-Podolsky-Rosen} steering inequalities from entropic uncertainty relations},
 journal={Phys. Rev. A}, volume={87}, pages={062103}, year={2013},
 doi={10.1103/PhysRevA.87.062103}}

@article{Berta10,
 author={Berta, Mario and Christandl, Matthias and Colbeck, Roger and Renes, Joseph M. and Renner, Renato},
 title={The uncertainty principle in the presence of quantum memory},
 journal={Nat. Phys.}, volume={6}, pages={659--662}, year={2010},
 doi={10.1038/nphys1734}}

@article{Seevinck07,
 author={Seevinck, Michael and Uffink, Jos},
 title={Local commutativity versus {Bell} inequality violation for entangled states and versus non-violation for separable states},
 journal={Phys. Rev. A}, volume={76}, pages={042105}, year={2007},
 doi={10.1103/PhysRevA.76.042105}}

@article{BuschLahtiWerner13,
 author={Busch, Paul and Lahti, Pekka and Werner, Reinhard F.},
 title={Proof of {Heisenberg's} error-disturbance relation},
 journal={Phys. Rev. Lett.}, volume={111}, pages={160405}, year={2013},
 doi={10.1103/PhysRevLett.111.160405}}

@article{Bardyn09,
 author={Bardyn, Charles-Edouard and Liew, Timothy C. H. and Massar, Serge and McKague, Matthew and Scarani, Valerio},
 title={Device-independent state estimation based on {Bell's} inequalities},
 journal={Phys. Rev. A}, volume={80}, pages={062327}, year={2009},
 doi={10.1103/PhysRevA.80.062327}}

@article{WuYuMolmer09,
 author={Wu, Shengjun and Yu, Sixia and M{\o}lmer, Klaus},
 title={Entropic uncertainty relation for mutually unbiased bases},
 journal={Phys. Rev. A}, volume={79}, pages={022104}, year={2009},
 doi={10.1103/PhysRevA.79.022104}}

@article{GisinBechmann98,
 author={Gisin, N. and Bechmann-Pasquinucci, H.},
 title={{Bell} inequality, {Bell} states and maximally entangled states for $n$ qubits},
 journal={Phys. Lett. A}, volume={246}, pages={1--6}, year={1998},
 doi={10.1016/S0375-9601(98)00516-7}}

@article{Bell64,
 author={Bell, J. S.},
 title={On the {Einstein Podolsky Rosen} paradox},
 journal={Physics Physique Fizika}, volume={1}, pages={195--200}, year={1964},
 doi={10.1103/PhysicsPhysiqueFizika.1.195}}

@article{LeggettGarg85,
 author={Leggett, A. J. and Garg, Anupam},
 title={Quantum mechanics versus macroscopic realism: Is the flux there when nobody looks?},
 journal={Phys. Rev. Lett.}, volume={54}, pages={857--860}, year={1985},
 doi={10.1103/PhysRevLett.54.857}}

@article{Coxeter71,
 author={Coxeter, H. S. M.},
 title={Frieze patterns},
 journal={Acta Arith.}, volume={18}, pages={297--310}, year={1971},
 doi={10.4064/aa-18-1-297-310}}

@article{Gunhan25,
 author={G{\"u}nhan, Ali Can and Gedik, Zafer},
 title={Information-theoretic and operational measures of quantum contextuality},
 journal={J. Phys. A: Math. Theor.}, volume={59}, pages={195302}, year={2026},
 doi={10.1088/1751-8121/ae67c0},
 eprint={2512.11049}, archivePrefix={arXiv}, primaryClass={quant-ph}}

@article{KoflerBrukner13,
 author={Kofler, Johannes and Brukner, {\v C}aslav},
 title={Condition for macroscopic realism beyond the {Leggett-Garg} inequalities},
 journal={Phys. Rev. A}, volume={87}, pages={052115}, year={2013},
 doi={10.1103/PhysRevA.87.052115}}

@article{Badziag11,
 author={Badzi{\k{a}}g, Piotr and Bengtsson, Ingemar and Cabello, Ad{\'a}n and Granstr{\"o}m, Helena and Larsson, Jan-{\AA}ke},
 title={Pentagrams and paradoxes},
 journal={Found. Phys.}, volume={41}, pages={414--423}, year={2011},
 doi={10.1007/s10701-010-9433-3}}

@article{CollinsGisin04,
 author={Collins, Daniel and Gisin, Nicolas},
 title={A relevant two qubit {Bell} inequality inequivalent to the {CHSH} inequality},
 journal={J. Phys. A}, volume={37}, pages={1775--1787}, year={2004},
 doi={10.1088/0305-4470/37/5/021}}

@article{BraunsteinCaves90,
 author={Braunstein, Samuel L. and Caves, Carlton M.},
 title={Wringing out better {Bell} inequalities},
 journal={Ann. Phys. (N.Y.)}, volume={202}, pages={22--56}, year={1990},
 doi={10.1016/0003-4916(90)90339-P}}

@misc{SM,
 note={See the Supplemental Material for verification code and figure data, available at \url{https://github.com/Affinewu/measurement_order}.}}

@article{CHSH69,
 author={Clauser, John F. and Horne, Michael A. and Shimony, Abner and Holt, Richard A.},
 title={Proposed experiment to test local hidden-variable theories},
 journal={Phys. Rev. Lett.}, volume={23}, pages={880--884}, year={1969},
 doi={10.1103/PhysRevLett.23.880}}
\end{document}